\documentclass{aa}
\usepackage{txfonts}
\usepackage{graphicx}
\begin{document}
 \title{The Enigmatic Young Object: Walker 90/V590 Monocerotis
   \thanks{Based on observations made with ESO Telescopes at Paranal 
           Observatory under programme ID 075.C-0528(A).}}

   \subtitle{}

   \author{M. R. P\'erez\inst{1}
          \and
          B. McCollum\inst{2}
          \and
	   M. E. van den Ancker\inst{3}
	    \and
	  M.D. Joner\inst {4}
	            }

   \offprints{M.R. P\'erez}

\institute{Los Alamos National Laboratory, P.O. Box 1663, ISR-1, MS B244, Los Alamos, NM 87545, U.S.A. \\
              \email{mperez@lanl.gov}
         \and
             Caltech, SIRTF Science Center, MS, 314-6, Pasadena, CA 91125, U.S.A.\\
	      \email{mccollum@ipac.caltech.edu} 
         \and
European Southern Observatory, Karl-Schwarzschild-Strasse 2, D-85748, Garching bei M\"unchen, Germany \\
             \email{mvandena@eso.org}        
        \and
             Brigham Young University, Dept. of Physics and Astronomy - ESC - N488, Provo, Utah 84602, U.S.A.\\ 
	     \email{jonerm@forty-two.byu.edu} 
             }

\date{Received April 2008; accepted}

\abstract{}{We assess the evolutionary status of the intriguing object Walker 90/V590 Mon, which is located about 20 arcminutes northwest of the Cone Nebula near the center of the open cluster NGC 2264. This object, according to its most recent optical spectral type determination (B7), which we confirmed, is at least 3 magnitudes too faint in $V$ for the cluster distance, but it shows the classical signs of a young pre-main sequence object, such as highly variable H$\alpha$ emission, Mg II emission, IR excess, UV continuum, and optical variability.}{We analyzed a collection of archival and original data on Walker 90, covering 45 years including photometry, imaging, and spectroscopic data ranging from ultraviolet to near-infrared wavelengths.}{According to star formation processes, it is expected that, as this object clears its primordial surroundings, it should become optically brighter, show a weakening of its IR excess and present decreasing line emissions. This behavior is supported by our observations and analysis, but timescales are expected to be longer than the one observed here. Based on photometric data secured in 2007, we find Walker 90 at its brightest recorded optical magnitude ($\overline{12.47} \pm 0.06$).  We document an evolution in spectral type over the past five decades (from A2/A3 to currently B7 and as early as B4), along with a decrease in the near-infrared K fluxes. From near-infrared {\it VISIR}~ images secured in 2004, Walker 90 appears as a point source placing an upper limit of $<0.1\arcsec$ for its diameter.  Evidence of turbulent inflows is found in rapidly changing inverse P-Cygni profiles in the lower Balmer lines, with a broadening of $\pm$ 400 km/s in H$\alpha$ and a redshifted component in H$\beta$ with a terminal velocity of $\sim$ 600 km/s. The measured steep UV continuum fluxes (mimicking a star as early as B4), added to a tentative identification of N V emission, suggest a strong non-photospheric component, typically of fluxes arising from a thermally inhomogeneous accretion disk.  We detect a well defined 2200\AA~ bump, indicative of dense material in the line-of-sight. We conclude that many observational features are explained if W90 is a flared disk system, surrounded by an inclined optically thick accretion disk.}{}	
\maketitle
%
\section{Introduction}

The immediate surroundings of young stellar objects are not well observed nor are they well understood. We have observed extensively the embedded object Walker 90 (V590 Mon, LkH$\alpha$ 25, NGC 2264-Vas62, HBC 219, IRAS 06379+0950, 2MASS J06404464+0948021, W90 hereafter), attracted by its puzzling behavior and uniqueness, with the goal of improving our understanding of the dynamics and constitution of the central object and its surroundings. Proper motion studies of the young open cluster NGC 2264 membership (e.g., Vasilevskis et al. 1965, Zhao et al. 1984) have determined that this object is a {\it bona fide} member with an estimated probability of membership of 92\% and 86\%, respectively. NGC 2264 is located in the Northern Monoceros region, with an age of 2--4 Myr (Lamm et al. 2004), and is an ideal area for probing primordial materials and by-products of star formation, such as dust, gas, CO, molecular clouds, neutral hydrogen and HII regions. In this cluster, as is also true in other star formation regions, young stars are often embedded and visual measurements are, therefore, challenging and somewhat unreliable.  The advantage of NGC 2264 over similar areas is that the foreground reddening is very small (E(B-V)=0.061) and well-established (Dahm et al. 2007), thereby eliminating foreground clumpiness as a source of uncertainty.  We present archival and new data on the enigmatic object W90, which is obscured by at least three magnitudes and is reddened by more than 0.2 magnitudes.  Because of these photometric properties, W90 has the rare privilege of being listed both as a T Tauri star (Herbig \& Rao 1972) and as a Herbig Ae/Be star (Herbig 1960, Finkenzeller \& Mundt 1984). Long-term observations and multiwavelength studies on this object will unequivocally define the dynamics of the local obscuration present as well as provide clues about the nature of the central source. 

This paper is organized as follows. The description of the multiwavelength new and unpublished data secured in the last two decades is found in Section 2. Subsequently, in Section 3 we discuss these data in the context of possible photometric and spectral variability and spectral type evolution. We also discuss in more detail the implications of diverse emission lines of interest that provide some insights into this enigmatic object.  Finally, in Section 4 we advance some conclusions of the current understanding of the physical conditions and evolutionary status of W90. 
\section{Observations}
\subsection{Optical and near-infrared photometry}
Photometric data have been secured by us for the last 20 years using different instruments. In subsequent sections we describe some of the most relevant results and setups used. In Tables 1 and 2, we present a selection of the original unpublished optical photometry. 
\begin{table}
\caption[]{Unpublished optical photometry of Walker 90 (from 1998 to 2005).}
\tabcolsep0.11cm
\begin{flushleft}
\begin{tabular}{lcccccl}
\hline\hline\noalign{\smallskip}
Date     & JD$-$2400000& $V$ & $B-V$& $V-R_c$ & $V-I_c$ & Obs.\\
\noalign{\smallskip}
\hline\noalign{\smallskip}
1988 Oct 31& 47466.311 & 12.651 & --&0.159 & 0.432 & WMO \\
1989 Jan 06& 47532.260 & 12.590 & --&0.197 & 0.472 & KPNO\\
1989 Jan 09& 47535.391 & 12.768 & --&0.143 & 0.326 & KPNO\\
1989 Jan 11& 47537.465 & 12.701 & --&0.142 & 0.321 & KPNO\\
1991 Jan 24& 48281.008 & 12.774 & --&0.158 & 0.304 & CTIO\\
1991 Jan 25& 48282.047 & 12.780 & --&0.187 & 0.381 & CTIO\\
1991 Jan 26& 48283.087 & 12.743 & --&0.177 & 0.379 & CTIO\\
1991 Jan 27& 48284.129 & 12.754 & --&0.178 & 0.352 & CTIO\\
1991 Mar 08& 48323.342 & 12.782 & --&0.171 & 0.336 & KPNO\\
1991 Mar 09& 48324.384 & 12.762 & --&0.175 & 0.339 & KPNO\\
1992 Mar 24& 48706.003 & 12.792 & --&0.186 & 0.422 & CTIO\\
1993 Dec 17& 49339.751 & 12.765 & --&0.258 & 0.323 & CTIO\\
2001 Mar 17& 51986.644 & 12.685 & --&--    & --    & BYU \\
2004 Jan 17& 53021.744 &12.859&0.132&0.148&0.377& TO \\
2004 Jan 27& 53031.715 &12.830&--&0.101&0.384& TO \\
2004 Feb 08& 53043.698 &12.806&0.191&0.115&0.336& TO \\
2004 Feb 13& 53048.687 &12.802&0.224&0.141&0.272& TO \\
2004 Feb 20& 53055.672 &12.758&0.204&0.028&0.333& TO \\
2004 Feb 26& 53061.757 &12.795&0.197&0.124&0.354& TO \\	
2004 Mar 09& 53073.714 &12.783&0.227&0.096&0.290& TO \\
2004 Mar 15& 53079.722 &12.753&0.245&0.083&0.328& TO \\
2004 Mar 18& 53082.724 &12.800&0.202&0.108&0.312& TO \\
2004 Mar 27& 53091.668 &12.771&0.213&0.082&0.327& TO \\
2004 Apr 09& 53104.698 &12.755&0.214&0.106&0.317& TO \\			
2004 Apr 13& 53108.679 &12.751&0.202&0.104&0.323& TO \\
2004 Apr 15& 53110.650 &12.772&0.178&0.100&0.322& TO \\
2005 May 06& 53496.644 &12.805&0.202&0.102&0.390& TO \\ 		
2005 Nov 22& 53696.908 &12.711&0.122&0.106&0.402& TO \\	
2005 Dec 05& 53709.992 &12.685&0.179&0.109&0.321& TO \\
2005 Dec 06& 53710.946 &12.689&0.199&0.096&0.239& TO \\
\noalign{\smallskip}
\hline
\end{tabular}
\end{flushleft}
\end{table}
\begin{table}
\caption[]{Unpublished optical photometry of Walker 90 (from 2006 to 2007).}
\tabcolsep0.11cm
\begin{flushleft}
\begin{tabular}{lcccccl}
\hline\hline\noalign{\smallskip}
Date     & JD$-$2400000& $V$ & $B-V$& $V-R_c$ & $V-I_c$ & Obs.\\
\noalign{\smallskip}
\hline\noalign{\smallskip}
2006 Jan 31& 53766.711 &12.717&0.141&0.107&0.364& TO \\
2006 Feb 04& 53770.717 &12.732&0.166&0.094&0.324& TO \\
2006 Feb 07& 53773.710 &12.712&0.193&0.104&0.309& TO \\				
2006 Mar 03& 53797.637 &12.719&0.186&0.108&0.328& TO \\
2006 Mar 18& 53812.610 &12.776&0.204&0.116&0.346& TO \\
2006 Mar 23& 53817.612 &12.790&0.215&0.120&0.373& TO \\
2006 Mar 26& 53820.627 &12.773&0.188&0.108&0.378& TO \\
2006 Apr 17& 53842.643 &12.784&0.238&0.099&0.347& TO \\
2006 Apr 18& 53843.666 &12.766&0.250&0.104&0.348& TO \\
2006 Apr 19& 53844.643 &12.752&0.238&0.123&0.354& TO \\		
2006 Oct 22& 54030.955 &12.632&0.203&0.088&0.289& TO \\
2006 Oct 23& 54031.950 &12.628&0.195&0.108&0.284& TO \\
2006 Oct 26& 54034.941 &12.666&0.179&0.068&0.369& TO \\	
2006 Nov 03& 54042.927 &12.623&0.166&0.099&0.290& TO \\
2006 Nov 07& 54046.905 &12.625&0.178&0.098&0.294& TO \\
2006 Nov 08& 54047.875 &12.626&0.201&0.072&0.251& TO \\	
2006 Nov 11& 54050.931 &12.629&0.211&0.074&0.227& TO \\
2006 Nov 12& 54051.933 &12.643&0.167&0.087&0.231& TO \\
2006 Nov 15& 54054.991 &12.616&0.186&0.117&0.282& TO \\
2006 Nov 16& 54055.988 &12.622&0.177&0.096&0.271& TO \\
2006 Nov 19& 54058.927 &12.622&0.183&0.078&0.219& TO \\
2006 Nov 20& 54059.919 &12.636&0.189&0.077&0.225& TO \\
2006 Nov 23& 54062.923 &12.591&-- &0.055&0.231& TO \\	
2006 Nov 24& 54063.913 &12.615&-- &0.089&0.262& TO \\		
2006 Nov 24& 54064.005 &12.613&-- &0.082&0.240& TO \\
2006 Nov 27& 54066.937 &12.635&-- &0.072&0.254& TO \\
2006 Nov 28& 54067.946 &12.621&0.212&094&0.187& TO \\		
2006 Dec 01& 54070.833 &12.605&-- &0.092&0.281& TO \\
2006 Dec 02& 54071.789 &12.612&-- &0.082&0.295& TO \\
2006 Dec 13& 54082.970 &12.601& 0.183&0.091&0.306& TO \\			
2006 Dec 14& 54083.965 &12.604&0.206&0.091&0.278& TO \\
2006 Dec 17& 54086.886 &12.616&0.188&0.086&0.281& TO \\
2006 Dec 21& 54090.829 &12.618&0.184&0.104&0.296& TO \\
2006 Dec 22& 54091.829 &12.641&--&0.102&0.337& TO \\
2006 Dec 25& 54094.842 &12.620&0.186&0.087&0.323& TO \\
2007 Jan 12& 54113.433 &12.524&0.124&--&--&SAAO \\
2007 Jan 13& 54114.420 &12.523&0.141&--&--&SAAO \\ 
2007 Jan 17& 54118.399 &12.394&0.129&--&--&SAAO \\
2007 Jan 18& 54119.375 &12.402&0.129&--&--&SAAO \\
2007 Jan 19& 54120.397 &12.396&0.134&--&--&SAAO \\
2007 Jan 20& 54121.367 &12.450&0.116&--&--&SAAO \\
2007 Jan 22& 54123.382 &12.543&0.110&--&--&SAAO \\
2007 Jan 23& 54124.361 &12.490&0.147&--&--&SAAO \\
2007 Feb 21& 54153.297 &12.479&0.135&--&--&SAAO \\
\noalign{\smallskip}
\hline
\end{tabular}
\end{flushleft}
\end{table}
We have used the new and unpublished photometric data with archival optical and near-infrared data taken from the literature (Walker 1956; Strom et al. 1971; Breger 1972; Breger 1974; Mendoza \& G\'omez 1980; Rydgren 1971; Sagar \& Joshi 1983; Kwon \& Lee 1983; P\'erez et al. 1987; Hillenbrand et al. 1992; Neri et al. 1993; Herbst \& Shevchenko 1999; de Winter et al. 2001; Sung et al. 1997, 2008; Lamm et al. 2004; Dahm \& Simon 2005) and near-IR photometry (Strom et al. 1972; Allen 1973; Warner et al. 1977; Sitko et al. 1984; Rydgren \& Vrba 1987; Neri et al. 1993; de Winter et al. 2001; 2MASS catalogue). We note that Sung et al. (2008) includes W90 under two names, W4792 and S2144, and list an extremely low $(B-V)$ value of 0.020 for observations made in 2002 January. Table 3 lists new near-infrared photometric measurements of W90. 
\begin{table}
\caption[]{Unpublished near-infrared photometry of Walker 90.}
\tabcolsep0.11cm
\begin{flushleft}
\begin{tabular}{lcccccl}
\hline\hline\noalign{\smallskip}
Date     & JD$-$2400000 & $J$ & $H$ & $K$ & $L$ & Obs.\\
\noalign{\smallskip}
\hline\noalign{\smallskip}
1984 Dec 01& 46036.0   & 11.49 & 10.71 & 9.23 & 7.57 & ESO \\
1990 Mar 16& 47966.5   & 11.48 & 10.45 & 9.31 & 7.44 & ESO \\
1990 Mar 17& 47967.5   & 11.57 & 10.49 & 9.31 & 7.55 & ESO \\
1990 Mar 18& 47968.5   & 11.55 & 10.49 & 9.33 & 7.48 & ESO \\
1990 Mar 19& 47969.5   & 11.53 & 10.48 & 9.31 & 7.36 & ESO \\
2003 Jan 29& 52669.3   & 11.20 & 10.16 & 9.34 & --   & SAAO\\
2003 Jan 30& 52692.3   & 11.86 & 10.90 & 9.71 & --   & SAAO\\
2004 Mar 24& 53089.0   & 10.99 & 10.11 & 9.24 & 7.58 & SAAO\\
2004 Mar 28& 53093.0   & 11.02 & 10.06 & 9.20 & --   & SAAO\\
\noalign{\smallskip}
\hline
\end{tabular}
\end{flushleft}
\end{table}

\subsection{H$\alpha$ index and $BVR_{c}I_{c}$ photometric data}

\begin{figure}
   \centering
 \includegraphics[angle=0, width=9.5cm]{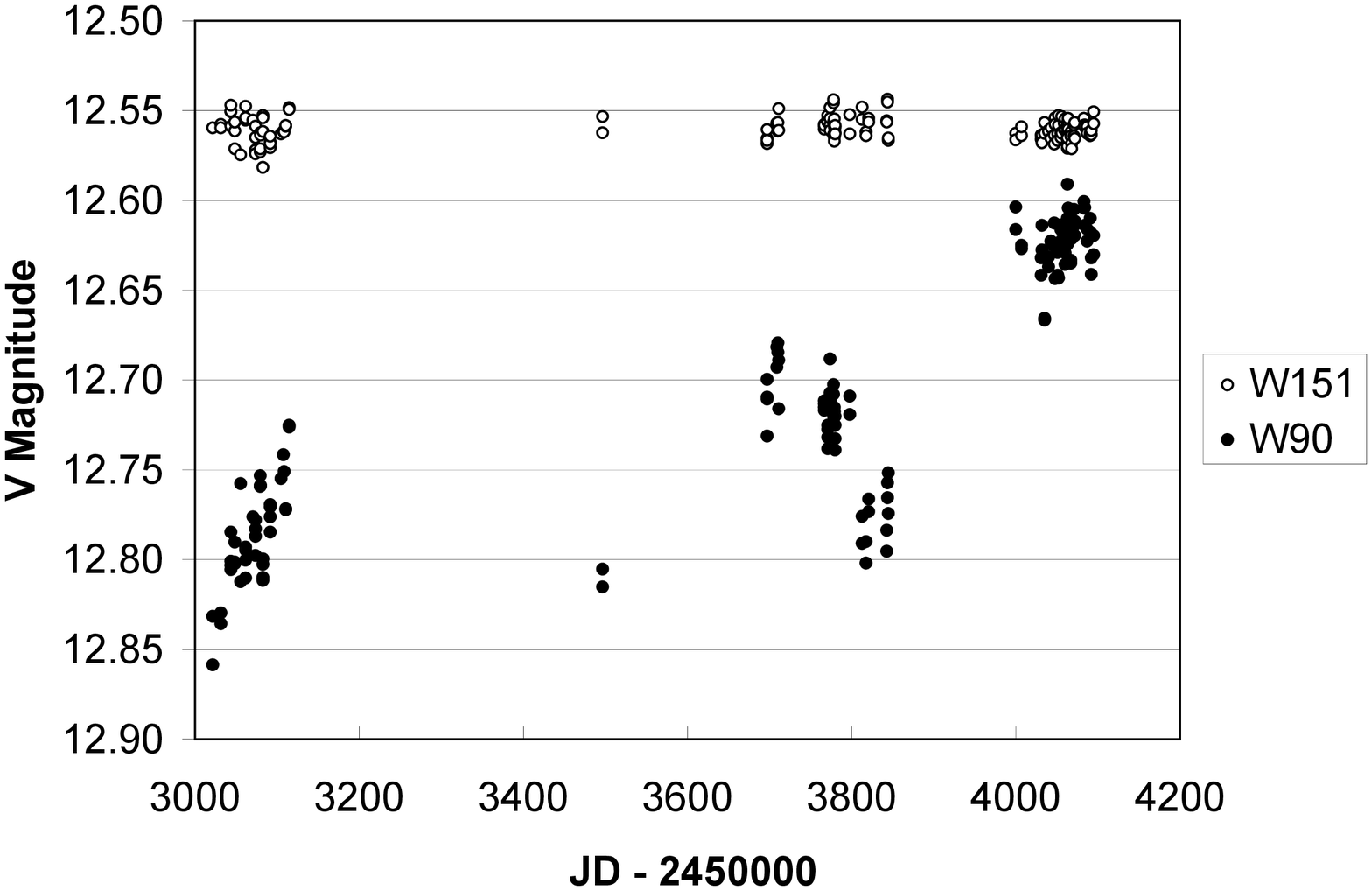}
      \caption{Photometric $V$ magnitude from 2004 January to 2006 December. The comparison star (W151, an older object) shows a $V$ magnitude around 12.55 during this period, whereas W90 appears brighter by about 0.20 magnitudes toward the end of the observing window.} 
         \label{Fig1}
   \end{figure}

\begin{figure}
   \centering
 \includegraphics[angle=0, width=9.5cm]{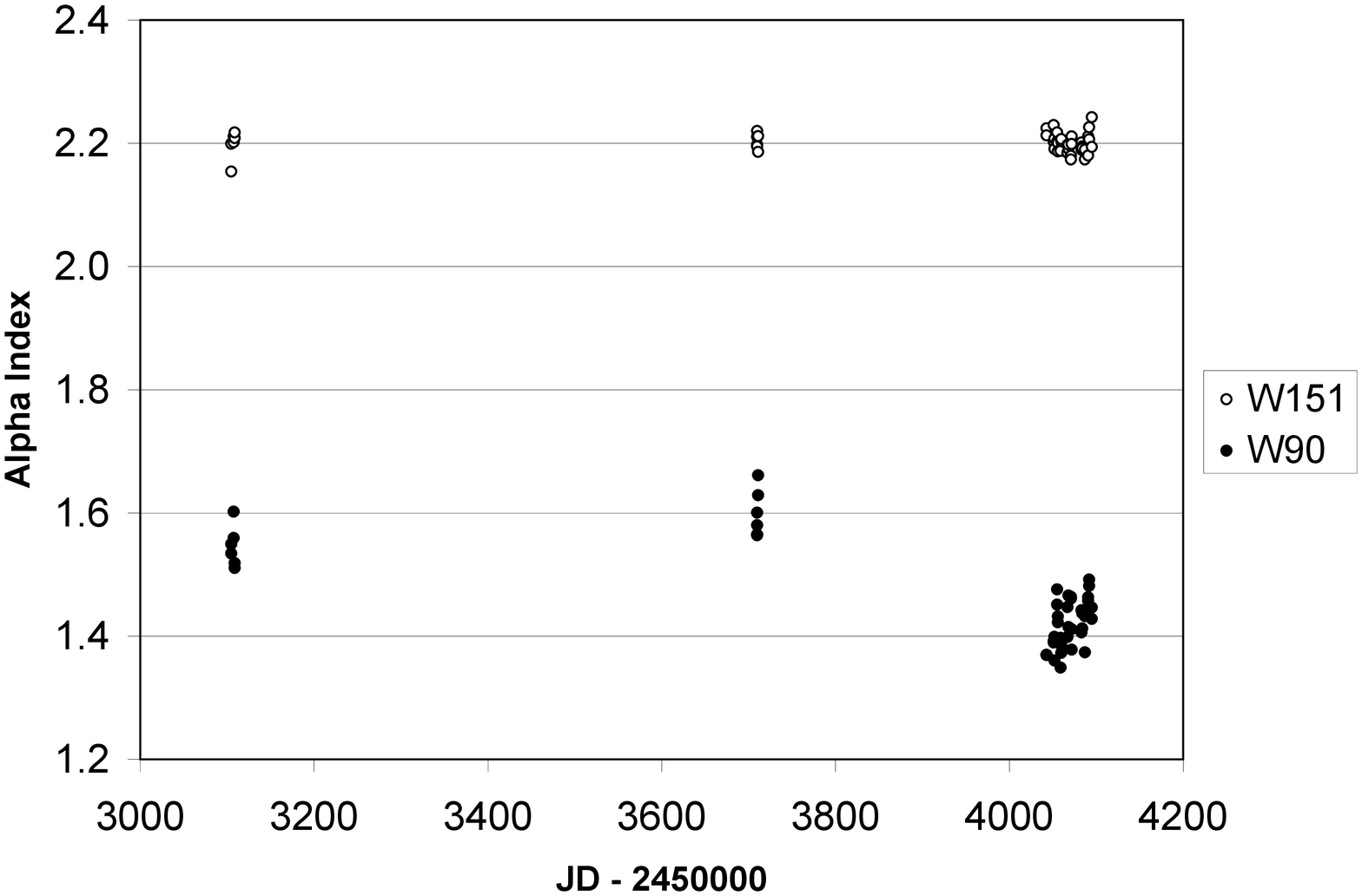}
      \caption{H$\alpha$ index from 2004 January to 2006 December. This index for W90 appears in emission throughout the observing window, especially toward the later times. The comparison star, W151, shows a constant index in absorption during this period.} 
         \label{Fig2}
   \end{figure}

The data shown in Figs.\ref{Fig1} and \ref{Fig2} and the majority of the new data displayed in Tables 1 and 2 were secured using the 0.81m robotic
telescope at the Tenagra Observatory (TO) in southern Arizona. The observations of 52 nights reported here, are part of a multi-year, multi-target monitoring program that was conducted at TO from 2004 January through 2006 December.


\subsubsection{$BVR_{c}I_{c}$ photometric data} Observations at TO were made using the back-illuminated SITe 1K CCD cooled to a temperature of -40$^\circ$C.  The plate scale for this system was found to be 0.87\arcsec per pixel. Filters were used to match the $BVR_{c}I_{c}$ Johnson/Cousins system and were patterened after the suggestions of Bessel (1979).  Calibration frames were taken each night to make bias, dark and flat field corrections to the program frames by using the routines found in IRAF.  Next, aperture photometry was done on each program frame using the APPHOT package in IRAF.  A fixed aperture with a 15.7\arcsec~ diameter was used for all the TO frames reported in this investigation.  Background subtraction was done using a sky annulus with a 3 pixel width surrounding the photometric aperture.

In addition to W90, aperture photometry was done on 11 other stars in the field.  From these, a comparison ensemble was selected that included five nearby stars (Walker numbers 93, 108, 117, 125, and 151) of similar magnitude that were observed every night.  Differential photometry was done between the comparison ensemble and W90 in order to obtain standard magnitude and color values for each of the observations through the various filters.  A sense for the variability of W90 can be noted from the fact that the dispersion per observation is about three times larger in $V$ than for nearby stars of similar magnitude.  Many nights had multiple observations and some time series photometry was done on several nights for this field.  Thus, the sample data presented in Tables 1 and 2 is a reduced set of all the data secured for W90 and as such represents average points of higher precision than single observations.

The most recent set of observations, covering 2007 January and February, was secured at SAAO, Sutherland, using the 0.5m telescope and the Modular Photometer (MP) with a GaAs photomultiplier and standard filters.  The transformation coefficients were carefully scrutinized by monitoring other field stars (Walker numbers 108, 125 and 151).  During a 40 day period covered by the observations, W90 was observed to be at the brightest $V$ magnitude of 12.39 and as faint as 12.54. Note that the dispersion per observation is 0.059 for W90 in $V$, which is also about three times as large as for stars of similar magnitude that were observed on the same nights. 
\subsubsection{H$\alpha$ Data} In addition, TO observations were made at three different epochs on a subset of the nights using a custom set of narrow (3nm) and wide (21nm) filters both centered on the H$\alpha$ line. The H$\alpha$ index is a wide/narrow photometric line index that is defined in the same manner as the H$\beta$ index described by Crawford \& Mander (1966). A system such as this is ideal for detecting stars with features in emission.  If the H$\alpha$ emission is strong, as is the case in W90, the index is unmistakably small. Individual H$\alpha$ spectral measurements in higher resolution showing their high variability are presented and discussed later in the paper. In this instrumental system, the ordinary main sequence stars with spectral types between A and G have an H$\alpha$ index that ranges between 2.3 and 2.0. Walker 151 (W151, GSC 00750-00997, V=12.58, spectral type=G2) was selected as a nearby comparison star because it had a main sequence H$\alpha$ value that was constant during the time of the observations. The W151 data show that another star of about the same magnitude as W90: 1) has an index around 2.2, as is to be expected for a star of this spectral type on the main sequence, and 2) has showed smaller variation overall at each observation epoch.  The H$\alpha$ lines in W90 were clearly in emission for all observations.  In fact, any star with an index smaller than about 1.8 is almost certainly showing emission in this system. As the index gets smaller after reaching the flat continuum value, increasingly stronger emission features are indicated for the last set of observations in 2006 December. 

\subsection{Optical spectroscopy}
\subsubsection{Low resolution data}

Low resolution spectroscopic data were secured at the ESO 1.5m telescope in three consecutive observing seasons between 1985 and 1987 with the Image Dissector Scanner (IDS). These data covered the spectral range from 3700 to 7000 \AA, with resolutions $\Delta\lambda \sim$ 1.8 and 0.6 \AA. Data reductions were accomplished to absolute flux.  For example, at the central wavelength of the $V$ filter (5500 \AA), the absolute flux is $2.7\times 10^{-13} watt m^{-2} \mu m^{-1}$, which corresponds to a $V$ magnitude of 12.90 by using the standard Johnson (1966) flux to magnitude conversion. 

\begin{figure}
   \centering
\includegraphics[angle=0,width=9.0cm]{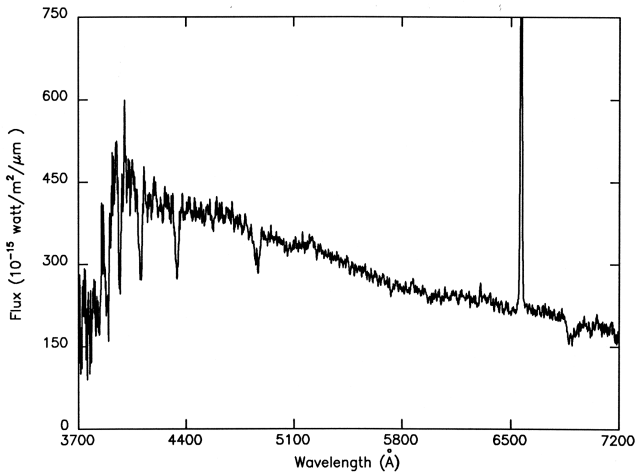}

      \caption{ Low resolution spectra ($\Delta \lambda \sim 1.8$ \AA) covering from 3700 to 7000 \AA. Note the strong flux continuum toward blue wavelengths. H$\alpha$ is in emission with an equivalent width of 10.5 \AA, H$\beta$ appears as a fill-in absorption, and other Balmer lines are in absorption.}
         \label{Fig3}
   \end{figure}

\begin{figure}
   \centering
\includegraphics[angle=0,width=9.0cm]{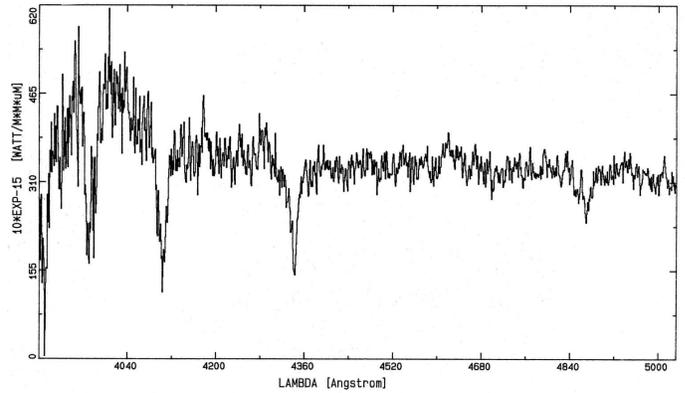}

      \caption{Low resolution spectra ($\Delta \lambda \sim 0.6$ \AA) centered at H$\gamma$. H$\beta$ appears barely in absorption with a narrow asymmetric emission superimposed. Other Balmer lines also appear asymmetric.}
         \label{Fig4}
   \end{figure}

In Fig.~\ref{Fig3} we present the observed spectrum of W90 covering over 3300 \AA. We note the strong flux continuum toward shorter wavelengths. Also H$\alpha$ is in emission, H$\beta$ appears in absorption with a filled-in emission component, and the remaining Balmer lines are in absorption. The flux distribution observed confirms the spectral type of B8pe at the time of the observations. Fig.~\ref{Fig4} illustrates in a medium resolution (R=$\lambda/\Delta \lambda=7,500$) the complex shape of the Balmer lines H$\beta$, H$\gamma$, H$\delta$ and H$\epsilon$. 

\subsubsection{Palomar Echelle data}

\begin{figure}
   \centering
 \includegraphics[width=8.5cm]{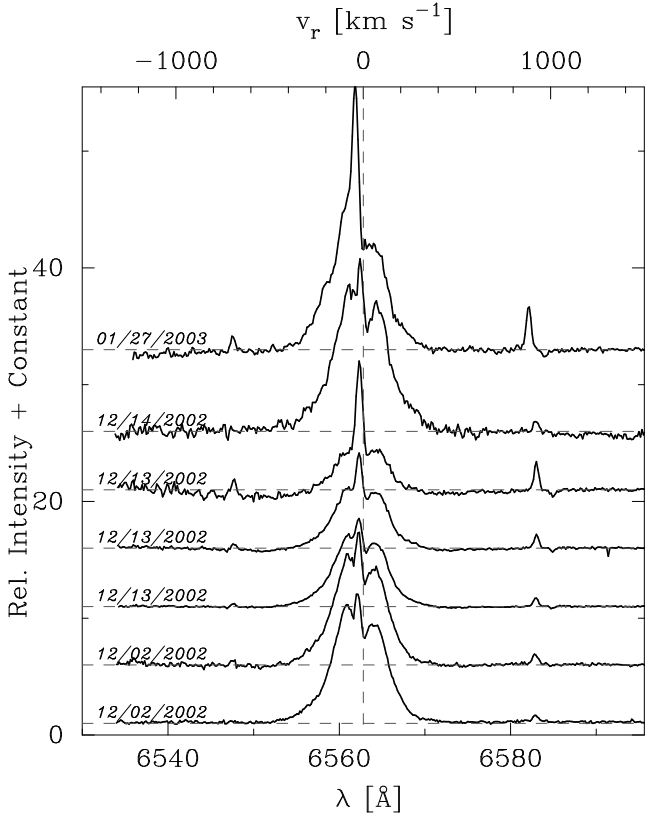}
      \caption{Seven observations of H$\alpha$ showing a highly variable structure, a broadening of $\pm$400 km/s, and a strong inverse P-Cygni profile on top of this broad emission. Note that in most cases $W_{H\alpha} \geq 10$~\AA. Observing dates are labeled.}
         \label{fig:halpha}
   \end{figure}

\begin{figure}
   \centering
 \includegraphics[width=8.5cm]{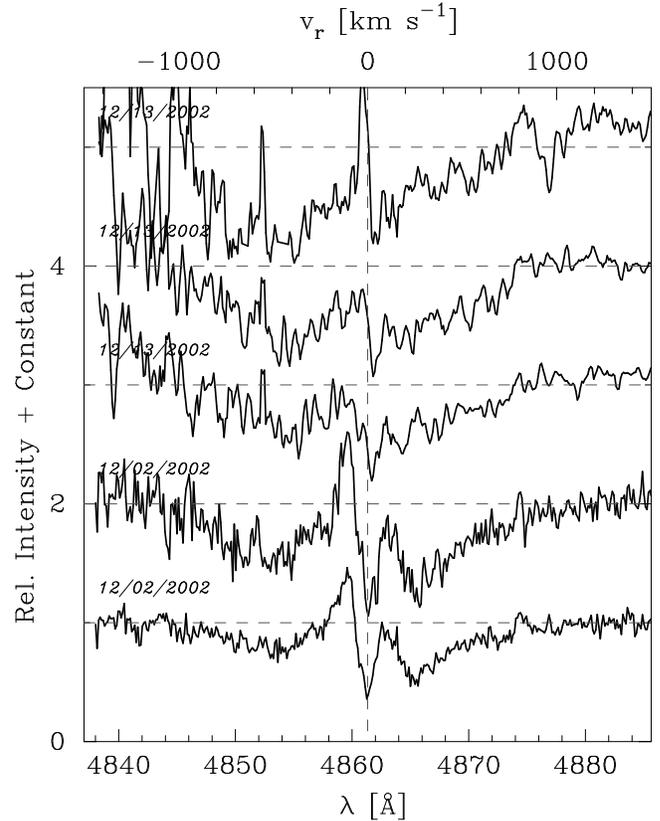}
      \caption{Multiple observations of H$\beta$ showing a highly variable strong inverse P-Cygni profile blueshifted and a redshifted component with terminal velocities of $\sim$ 500 and $\sim$ 600 km/s, respectively.}
         \label{fig:hbeta}
   \end{figure}

\begin{figure}
   \centering
 \includegraphics[width=8.5cm]{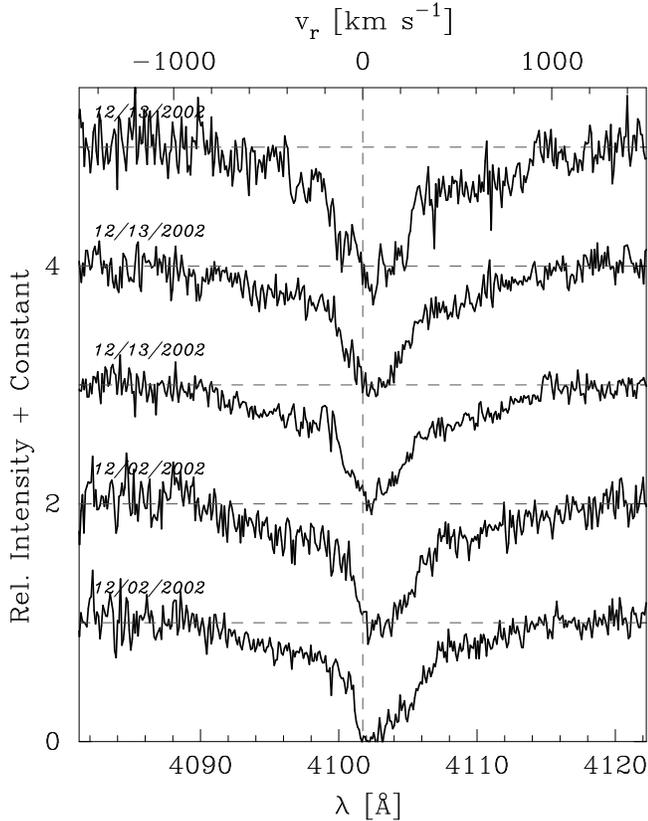}
      \caption{Five observations of H$\delta$ showing a highly variable profile and strong redshifted asymmetric profile with a terminal velocity of $\sim$ 400 km/s. This indicates the presence of circumstellar material moving away in our line-of-sight, which could be due to stellar winds or to neutral gas accretion flows typical in Herbig Ae/Be Stars (Ghandour et al. 1994). The same dynamics are present in the profiles for H$\delta$ and H$\epsilon$.}
         \label{fig:hdelta}
   \end{figure}

\begin{figure}
   \centering
 \includegraphics[width=8.5cm]{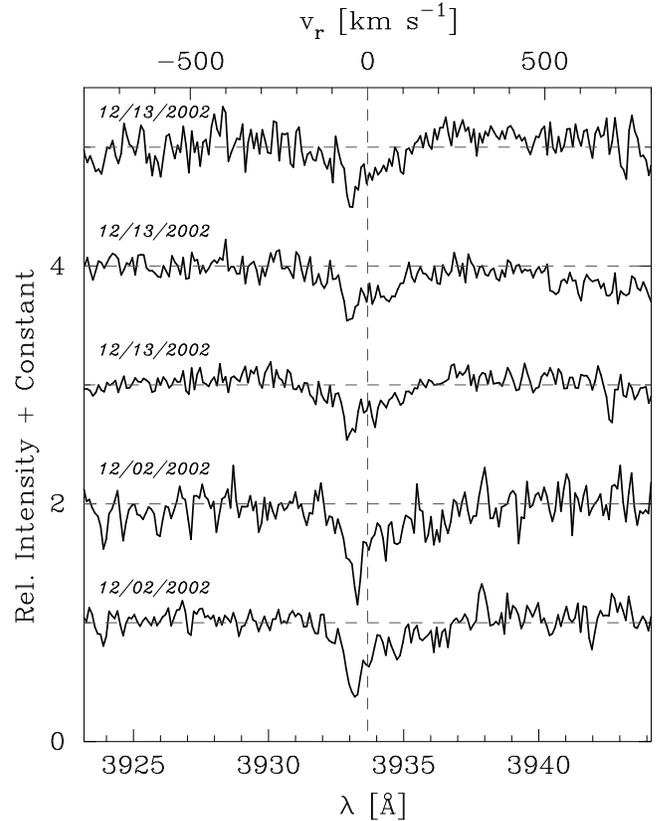}
      \caption{Observations of Ca II K line (3934 \AA) showing a variable profile in weak absorption. Note the modest blueshifted centroid of this line.}
         \label{fig:caiik}
   \end{figure}

Optical spectra were obtained at the Mt. Palomar 60-inch telescope on 14 observing nights during the period from 2002 November through 2003 March. We used the Norris Echelle spectrograph, which provided a resolution of R = 19,000 ($\Delta\lambda$ = 0.2-0.5~\AA) and covered the wavelength range from 3600 to 9000~\AA.  We used the CCD9, a Texas Instruments backside-illuminated 800 x 800 CCD, along with a 2.13 x 7.90\arcsec slit.  Most exposures have S/N ratios measured at 6,000 \AA, ranging between 10 to 30, with an average of 18. Table 4 lists all the exposures taken and their duration. Photometric conditions were not ideal and the targets were often partially obscured by cloud cover which sometimes produced variable obscuration during a single exposure.  Seeing varied from $\sim$1 - 3\arcsec, with a mean around 1.5\arcsec to 2\arcsec.  The spectrograph was manually rotated to within about a degree of the parallactic axis at the beginning of most observations, and could not be changed during the course of an exposure. 
\begin{table}
\caption{W90 optical spectra obtained with Palomar 60-inch}             
\label{table:4}      
\centering                          
\begin{tabular}{c c c c c}        
\hline\hline                 
 Date & Exp. Start& JD$-$2400000&Duration\\    
& (UT)& &(min) \\
\hline                        
   2002 Nov 16 & 07:06  & 52594.79&70\\
   2002 Nov 16 & 09:40  & 52594.90&120\\
   2002 Nov 17 & 08:19  & 52595.84&90\\
   2002 Dec 02 & 06:48  & 52610.78&60\\
   2002 Dec 02 & 08:08  & 52610.83&60\\
   2002 Dec 13 & 05:45  & 52621.73&60\\
   2002 Dec 13 & 07:06  & 52621.79&60\\
   2002 Dec 13 & 08:31  & 52621.85&60\\
   2003 Dec 14 & 12:14  & 52623.01&32\\
   2003 Jan 10 & 10:20  & 52649.93&60\\
   2003 Jan 11 & 08:45  & 52650.86&60\\
   2003 Jan 19 & 08:59  & 52658.87&30\\
   2003 Jan 19 & 09:51  & 52658.91&50\\
   2003 Jan 19 & 10:55  & 52658.95&50\\
   2003 Jan 19 & 12:00  & 52659.00&25\\
   2003 Jan 27 & 06:07  & 52666.75&60\\
   2003 Jan 27 & 10:25  & 52666.93&50\\
   2003 Jan 27 & 11:16  & 52666.96&40\\
   2003 Feb 18 & 06:44  & 52688.78&60\\
   2003 Feb 18 & 07:51  & 52688.82&60\\
   2003 Feb 18 & 09:00  & 52688.87&60\\
   2003 Feb 19 & 03:51  & 52689.66&60\\
   2003 Feb 19 & 08:23  & 52689.84&60\\
   2003 Mar 14 & 03:30  & 52712.64&60\\
   2003 Mar 14 & 04:30  & 52712.68&60\\
   2003 Mar 15 & 02:53  & 52713.62&25\\
\hline                                   
\end{tabular}
\end{table}
A total of 26 Echelle spectra of W90 and of the comparison star, Walker 212 (HD 47961, B2V, V=7.5), were secured. The spectral variability, both in continuum flux and lines, presents changes on a time scale as small as hours. This is not uncommon in HAeBe stars; however, the cause of these variabilities remains unclear. In Figs.\ref{fig:halpha}--\ref{fig:caiik}, we present the Echelle line spectra of W90. A discussion of the specific lines observed follows in Section 3. 

\subsection{Ultraviolet spectroscopy}

The {\it International Ultraviolet Explorer (IUE)} archives contain 17 low-dispersion ($\Delta\lambda \sim$ 6~\AA) spectra of W90, ten long-wavelength (LWP), and seven short-wavelength (SWP) camera spectra, all obtained through the large aperture (10\arcsec\ $\times$ 20\arcsec). These spectra were secured between 1980 and 1994. Since the optimal exposure times for each wavelength region are about 6-8 hours long and because the largest exposure time was only 180 minutes long, the S/N in the data presented here are low and the spectra are generally underexposed. 

By using the Fine Error Sensor (FES) counts recorded on the IUE observing scripts and the photometric calibration by P\'erez \& Loomis (1991), we estimated $V$ magnitudes for the time of the IUE exposures for all spectra with the exception of LWP 27428. This $V(FES)$ calibration has a mean error of $\pm$ 0.03 mag. FES counts were affected by the abnormal increase in scattered light seen by the telescope optics after 1991 January 22. We have noted the $V$ estimates after 1991 with an asterisk in Table 5, fourth column, indicating that they are more uncertain than the values from previous years and quite likely overestimate the actual $V$ magnitudes. 
\begin{table}
\caption[]{{\it IUE} short- and long-wavelength camera data for W90.}
\tabcolsep0.11cm
\begin{flushleft}
\begin{tabular}{lccll}
\hline\hline\noalign{\smallskip}
Date     & Image.&Exp. Time&~$V(FES)$&Comments\\
&Number&(min)&(mag)& \\
\noalign{\smallskip}
\hline\noalign{\smallskip}
1980 Feb 02& SWP07991& 150&12.54& \\
1980 Nov 23& SWP10661& 120&12.58&underexposed\\
1981 Jan 14& SWP11046& 150&12.59& \\
1982 Jan 11& SWP16016& 120&12.50&high background\\
1992 Feb 18& SWP44015& 180&12.82*& \\
1992 Feb 20& SWP44024& 130&13.03*&\\
1992 Mar 28& SWP44261& 173&13.07*&\\\hline

1980 Feb 19& LWR06956& 90&12.57&underexposed\\
1980 Feb 22& LWR06970& 85&12.62&underexposed\\
1980 Nov 23& LWR09369& 105&12.56& \\
1981 Jan 13& LWR09701& 120&12.63& \\
1982 Jan 11& LWR12324& 105&12.49& \\
1992 Feb 17& LWP22408& 120&12.80*& \\
1992 Feb 20& LWP22417& 90&13.10*&underexposed\\
1992 Mar 28& LWP22701& 80&13.08*&underexposed\\
1992 Mar 28& LWP22702& 90&12.98*&underexposed\\
1994 Feb 16& LWP27428& 80&--&abnormal fluxes\\
\noalign{\smallskip}
\hline
\end{tabular}
\end{flushleft}
\end{table}

\subsubsection{Short wavelength ultraviolet spectra}

\begin{figure}
   \centering
 \includegraphics[width=8.5cm]{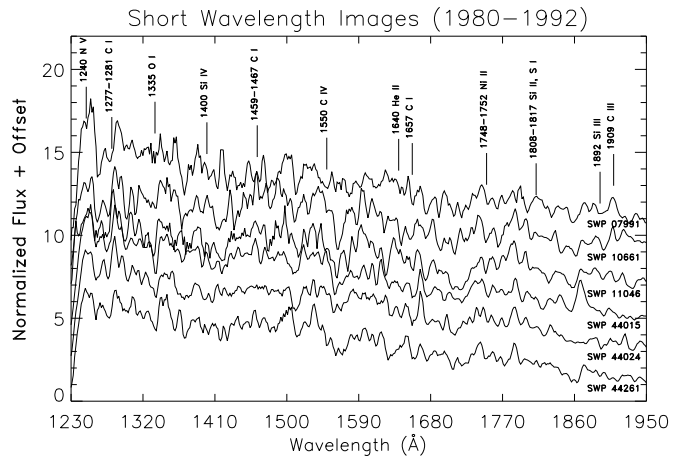}
      \caption{Archive SW spectra.  A sample of the spectra available with the best S/N (excluding SWP 16016). We present line tentative identification and note the variabilities in the continuum and line fluxes.}
         \label{fig:swp}
   \end{figure}

\begin{figure}
   \centering
 \includegraphics[width=8.5cm]{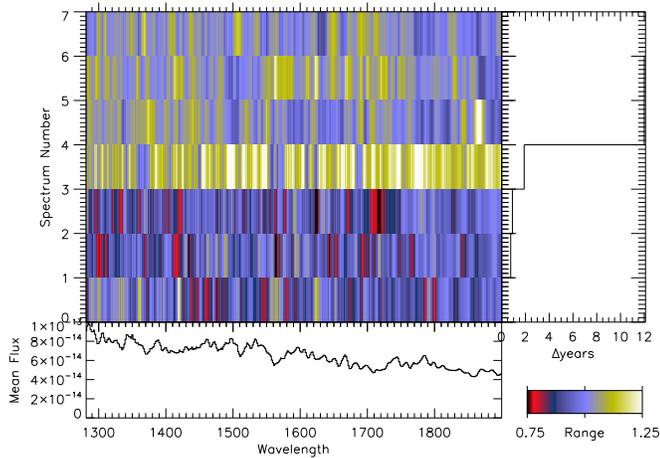}
      \caption{Short wavelength spectra. The seven spectra available for W90 ordered by date are stacked up in this color figure. The bottom insert includes a co-added flux spectrum of the seven individual NEWSIPS spectra, flux corrected by the Massa \& Fitzpatrick (2000) calibration. Note the clear continuum rising toward the short wavelength region.}
         \label{fig:swp_color}
   \end{figure}

In Fig.~\ref{fig:swp} we included a mosaic of the six short wavelength spectra (1230 to 1950 \AA) taken during a span of 12 years (excluding SWP 16016). We note the variabilities in the continuum and in the emission lines indicated. The flux data include the NEWSIPS flux calibration.  Fig.~\ref{fig:swp_color} presents all seven SWP spectra stacked up by date (from the bottom up).  In the bottom insert, a co-added flux spectrum is depicted. This shows some of the increasing continuum toward shorter wavelengths. We point out that image 4 (SWP 16016) appears with a strong continuum flux, possibly due to background radiation which saturates portions of the camera. We also note a less dramatic brightening in image 6 (SWP 44024) with a marginally larger exposure time of 130 minutes. We note that because of the length of the exposures, Lyman $\alpha$ geo-coronal emission (1215 \AA) strongly contaminated all short-wavelength spectra. As a result, it is impossible to determine whether W90 has an intrinsic emission at this wavelength. That saturated part of the spectra was excluded from Figs.~\ref{fig:swp} and \ref{fig:swp_color}.  

\subsubsection{Long wavelength ultraviolet spectra}

\begin{figure}
   \centering
 \includegraphics[width=8.5cm]{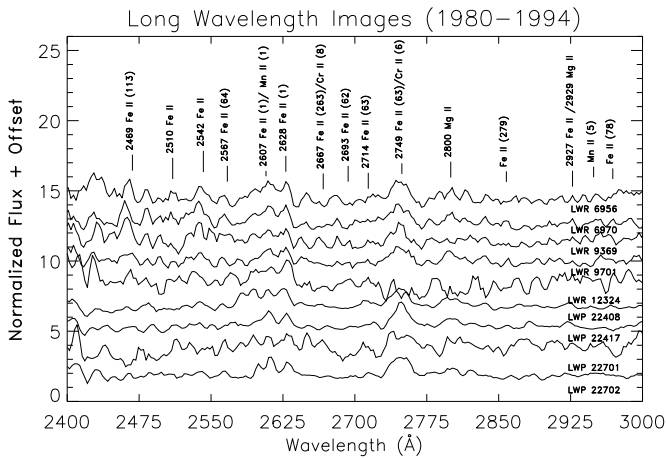}
      \caption{Archive LW spectra. A sample of the spectra available with the best S/N (excluding LWP 27428). Note the variability in the Fe I, Fe II, and Mg II emission lines.}
         \label{fig:lwp}
   \end{figure}

\begin{figure}
   \centering
 \includegraphics[width=8.5cm]{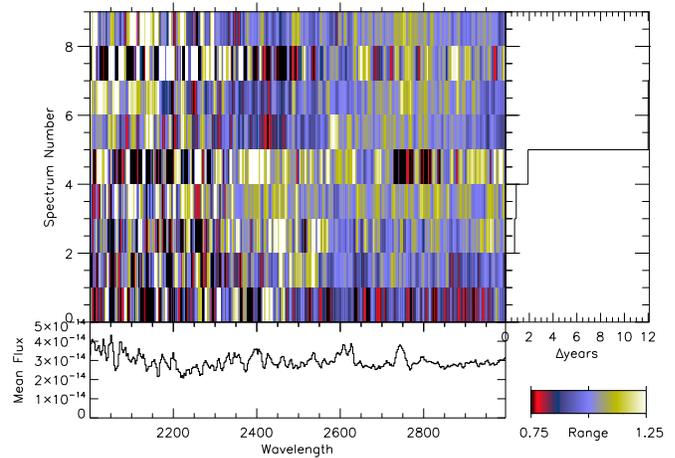}
      \caption{Long wavelength spectra. The ten spectra available for W90 ordered by date are stacked up in this color figure. The bottom insert includes a co-added flux spectrum of the ten individual NEWSIPS spectra, flux corrected by the Massa \& Fitzpatrick (2000) calibration. Many emission lines are presented, primarily Fe I, Fe II, and Mg II.}
         \label{fig:lwp_color}
   \end{figure}

In Fig.~\ref{fig:lwp} we present nine long wavelength spectra (2400--3000 \AA) of exposure times ranging from 90 to 120 minutes taken over a span of 14 years.  The striking variability of the overall flux spectrum, along with the strongest lines, can be clearly seen. Other variability (especially at $\lambda \leq 2450 \AA$) could be an artifact due to the low S/N of the spectra. Fig.~\ref{fig:lwp_color} presents all ten LWP spectra available stacked up by date (from the bottom up). In the bottom insert, a co-added flux spectrum is depicted showing some well developed emission lines and the 2200 \AA~ broad absorption feature due to interstellar extinction, which probes a dense line-of-sight with some possible large organic molecules. We note that the 2200 \AA~ absorption feature, although common in galaxies and other distant objects, is rarely detected in galactic stellar objects such as W90, which clearly suggests high density of small particles in the line-of-sight (Steenman \& Th\'{e} 1991). These data were reduced by applying the NEWSIPS low-dispersion flux correction devised by Massa \& Fitzpatrick (2000), which corrected flux errors of $10-15$\%. 

\subsubsection{Infrared imaging: {\it Spitzer} and VLT}

We examined {\it Spitzer} archival IRAC (Fazio et al. 2004) and MIPS (Rieke et al. 2004) images of W90 and its environs.  The IRAC data, which have a resolution of 1.25\arcsec per pixel, show W90 as an unresolved source in all four filters (3.6, 4, 5, 5.8, and 8.0 $\mu$m).  The MIPS images have resolutions of 2.55\arcsec per pixel at 24 $\mu$m, and 4.99\arcsec per pixel at 70 $\mu$m.  At 24 $\mu$m W90 is unresolved, while at 70 $\mu$m it is not distinguishable as a discrete source amid the bright extended emission at that location.  Considerable extended emission is apparent within $\sim$5\arcmin or more in all directions from W90 at all {\it Spitzer} wavelengths, but the emission appears to be typical ISM background and does not show any structure which suggests a physical relation to W90. 

Diffraction-limited images of W90 in the 11.3 and 11.9~$\mu$m bands were also obtained with the VLT Imager and Spectrometer for the mid-Infrared ({\it VISIR}; Lagage et al. 2004) at ESO-Paranal during the night of 2004 December 12. Subtraction of the thermal emission from the sky, as well as the telescope itself, was achieved by chopping in the North-South direction with a chop throw of 15\arcsec, and nodding the telescope in the opposite direction with equal amplitude.  The cosmetic quality of the images was further improved by superimposing a random jitter pattern (with a maximum throw 2\arcsec) on the nodding sequence, so as to minimize the effect of bad pixels in the detector array on the final science data.  Total integration times were 13 and 18 minutes for the 11.3 and the 11.9~$\mu$m filters, respectively.

\section{Discussion and results}
\subsection{Cluster membership, photometric variability and evolution}

Object W90 is likely to be a cluster member due to its physical location, proper motion, spectral type, and particular reddening characteristics, observables which have contributed to a high probability of membership ($\sim$ 90\%), as indicated earlier. Since Herbig (1954) classified this star as entry No. 25 (LkH$\alpha$ 25) with a photographic magnitude of 14.0 in his catalog of emission-line objects, increasing interest has been drawn to this peculiar star. Later, Walker (1956) with photoelectric measurements found magnitude variations from 13.18 to 12.88, and thus classified it as a potential variable. Now it is also known as V590 Mon (Herbig and Bell 1988). Likewise, Walker (1956) also discovered that this star lies below the ZAMS in the H-R diagram because of an unusual reddening in excess of 3 to 4 magnitudes. Sitko et al. (1984) noted that W90 was 1 magnitude below the ZAMS and about 3 magnitudes below the location of other cluster members in NGC 2264. W90 location on the H-R diagram is hardly unique, since recently Sung et al. (2008) identified 82 below main sequence (BMS) objects brighter than $I_c=20$ magnitudes in NGC 2264, including W90. Because of their near-IR excesses, most of these BMS stars appear to be pre-main sequence objects seen through circumstellar material. 

\begin{figure}
   \centering
 \includegraphics[angle=0,width=8.5cm]{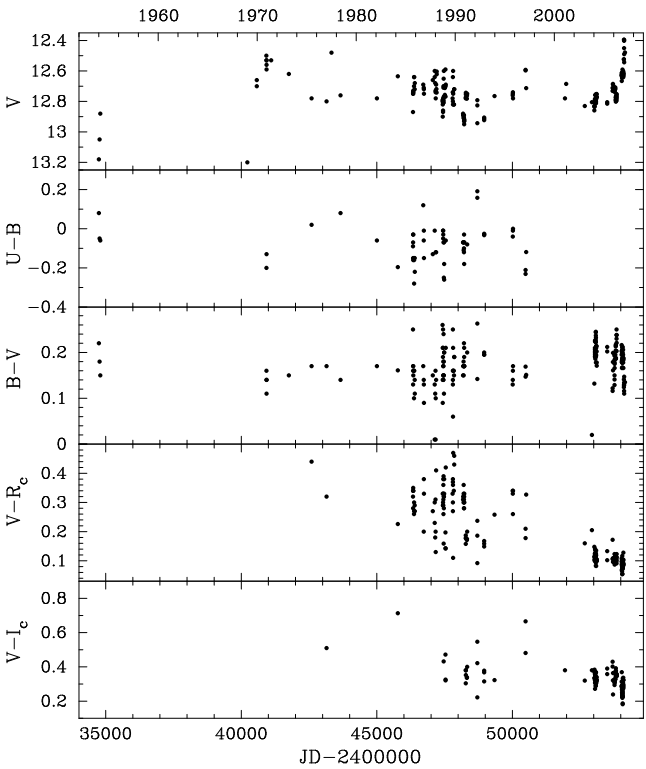}
      \caption{Visual magnitude and color variabilities for the last 45 years. A quick inspection reveals the lack of correlation among the color changes and the $V$ magnitude. After 2002, W90 seems to be at its brightest level with dramatic color changes in the indices $V-R_c$ and $V-I_c$.} 
         \label{fig:45 years}
   \end{figure}

\begin{figure}
   \centering
 \includegraphics[angle=-90,width=8.5cm]{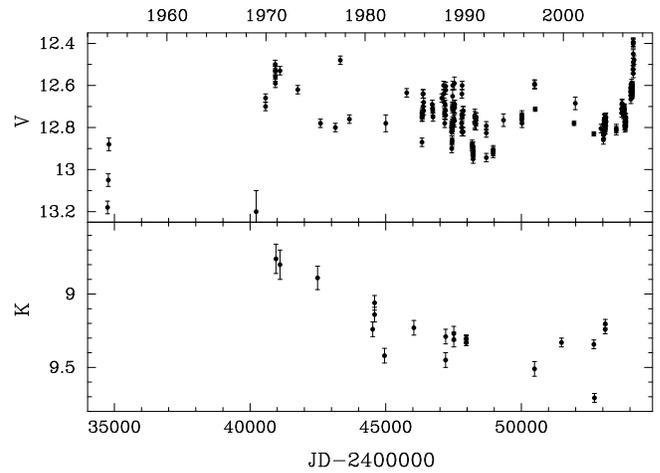}
      \caption{Variability of $V$ and $K$ magnitudes for the last 45 years. Note the systematic decrease of the $K$ color flux in the last 30 years. Modeling this decrease as $\Delta(K) \sim -\alpha(t-t_{o})$, $\alpha$ is 0.22 magnitudes per decade, or a decrease in flux, $F_{K}$, by 42\% in the last three decades.} 
         \label{fig:long-term var}
   \end{figure}

For example, if this obscuration in W90 is caused by the remnant primordial material surrounding this object, an eventual clearance process is expected, yielding increased optical fluxes.  This process was thought to have occurred when Bhatt \& Sagar (1992), using observations taken in 1989, announced that this object was 3 magnitudes brighter at V=9.7. Unfortunately, this was demonstrated to be a false alarm (P\'erez et al. 1992). Our observations for the last decade show small variabilities, V$=12.68 \pm 0.11$, with some brightening trends, as is shown in the following section. We present in Fig.~\ref{fig:45 years} a compilation of 45 years of optical Johnson and Cousins photometric values taken from the literature and from our unpublished data. Trends apparent in the data can be better determined in subsequent figures. 
We present further evidence of the primary decrease in brightness in the $K$ color filter for this star. In Fig.~\ref{fig:long-term var} we illustrate the apparent increase of the luminosity in the visual magnitude, $V$, since 1956 (we excluded Herbig's photographic magnitude) and the sustained luminosity decrease of the near-infrared magnitude, $K$, over the same period. One can note the systematic weakening of the $K$ magnitudes in the recent past, of 0.22 magnitudes per decade, whereas the $V$ magnitude seems to have non-periodic faint and bright incursions. 

This brightening effect can be attributed to either a change in the inclination 
angle, i, or to a real vanishing, by expulsion or accretion, of the thick 
circumstellar shell around this star. Rydgren and Vrba (1987) argued that the large infrared excess of this star can be explained as an optical depression, and that this abnormal extinction can be attributed to an edge-on circumstellar disk.

\begin{figure}
   \centering
 \includegraphics[angle=-90,width=8.5cm]{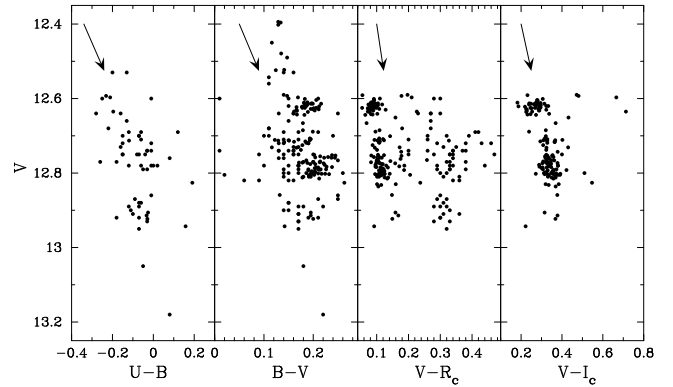}
      \caption{Color variations for the last 45 years. The interstellar reddening line is indicated with an arrow in each panel. Note manifestations of the blueing effect in the $V$ versus $(U-B)$ and $(V-I_c)$ diagrams. As the star becomes fainter on some occasions, instead of becoming redder the $(B-V)$ colors remain at the same color or they become slightly bluer. A similar effect is presented in the other panels.}
         \label{fig:color var}
   \end{figure}

\begin{figure}
   \centering
 \includegraphics[angle=-90,width=8.5cm]{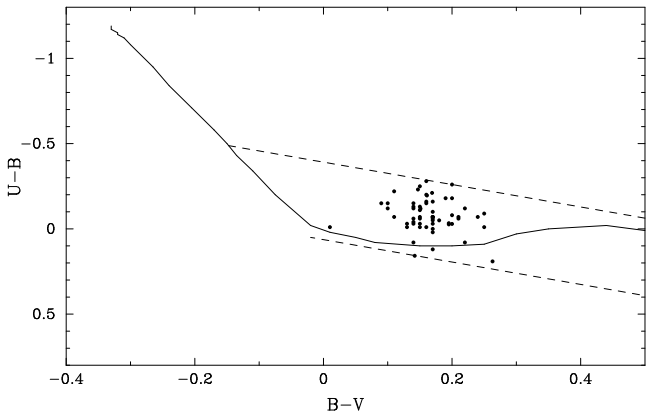}
      \caption{Observed color-color diagram, $(U-B)$ against $(B-V)$ for the archival and new data. Note the grouping of the data points between the dashed lines. The solid line represents the ZAMS. W90 data points are located mostly above the flat part of the ZAMS line, mimicking a somewhat later spectral type.}
         \label{fig:cc_UBV}
   \end{figure}

In Fig.~\ref{fig:color var} we illustrate the visual magnitude against the Johnson/Cousins color variations for the same data set secured in the last 45 years. The clustering of the data seems to depart in each panel from the interstellar reddening line, indicated with an arrow. The most noticeable cases are the $(U-B)$ and $(V-I_c)$ data sets illustrating an incipient blueing effect (``hockey stick'' shape), or Algol type brightness minima, which is a well-known phenomenon present among UX Ori objects (e.g., Grinin et al. 1994).    
In the context of the more well-studied variabilities in T Tauri stars (e.g, Herbst et al. 1994), one can test the hypothesis of whether the color variabilities of W90 follow the ortho-UBV or para-UBV patterns proposed by Safier (1995).  Fig.~\ref{fig:cc_UBV} presents the color-color diagram of the available data points. Neither classification scheme devised by Safier (1995) is obviously evident. However, by simple elimination, W90 is less of an ortho-UBV and possibly more of a para-UBV object (i.e., $U-B$ colors tend to be anticorrelated with $B-V$). No clear or unique explanation exists for the variability of the para-UBV objects in T Tauri stars, but Safier (1995) has speculated that these changes could correspond to magnetic accretion, provided that the continuum emission arises in the accretion shock at the stellar surface with fixed disk density and a variable accretion radius. The accretion disk interpretation in W90 is further discussed in later sections.   

\begin{figure}
   \centering
 \includegraphics[angle=-90,width=8.5cm]{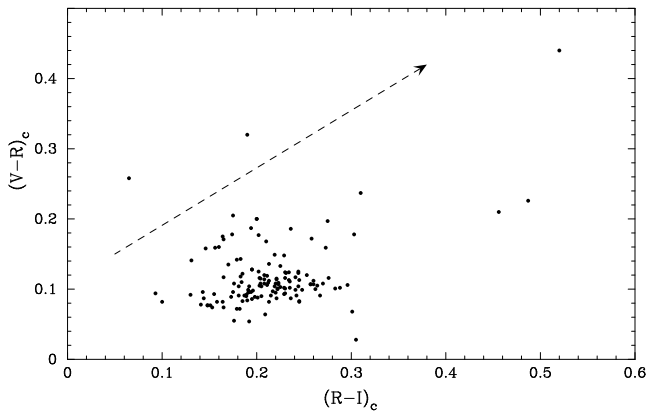}
      \caption{Observed color-color diagram $(V-R)_c$ against $(R-I)_c$ for the archival and new data. On this plane, the data points follow the trend outlined by the dashed line, which represents the reddening vector. The variability at longer wavelengths seems dominated by extinction changes along the line-of-sight.} 
         \label{fig:cc_VRI}
   \end{figure}

We have also plotted the available photometric data at longer wavelengths in the plane of $(V-R)_c$ versus $(R-I)_c$ in Fig.~\ref{fig:cc_VRI}. The data points clearly follow the reddening vector represented by the dashed line. In the classification scheme presented by Safier (1995), this resembles the ortho-UBV in T Tauri stars, which could be explained as variable obscuration as seems to be the case for the $(V-R)_c$ versus $(R-I)_c$ colors.  However, we warn that the plane of $VRI$ variability remains unexplored in the context of Safier's (1995) classification scheme. 

\subsection{Evolution of spectral type}

It is interesting to note the apparent evolution toward earlier spectral
types for this star. Herbig (1954) classified it as A2-A3; then Walker
(1956) gave it a spectral type of A2p, Herbig (1960) B8pe+shell, Warner
et al. (1977), B9-A0, and Young (1978), B4V. Nevertheless, P\'{e}rez et al. (1987) found a spectral type of B8V from IDS data, although the derived photometric spectral type is B9.5V (Q method), suggesting a discrepancy 
between its spectroscopic and photometric behavior.  Based purely on the strong UV fluxes toward short wavelengths, Imhoff \& Appenzeller (1987) classified W90 as B4. The most recent determination by Hillenbrand (1995) assigned a B7 spectral type to W90. Hern\'andez et al. (2004) used new spectroscopic data and reviewed some of the published spectral type classifications before concluding that the spectral type was B7 with a 2.0 spectral subtype error.  

\subsection{What spectral type is it?}

\begin{figure}
   \centering
 \includegraphics[angle=-90,width=8.5cm]{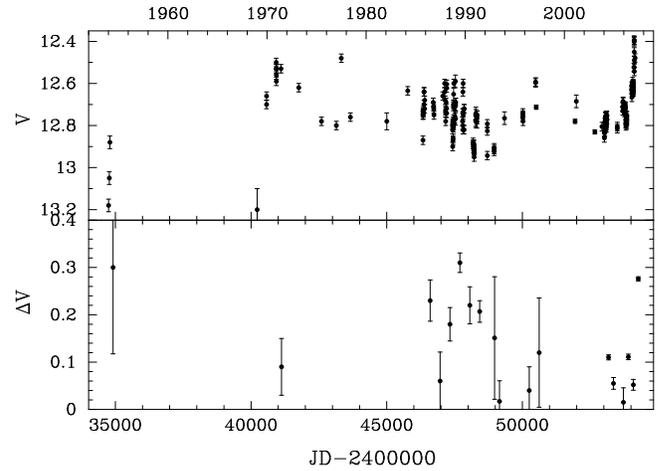}
      \caption{Variability of $V$ and $\Delta V$ against time. $\Delta V$ was formed by subtracting $V_{min}$ from $V_{max}$ for each year with multiple observations. Note that after the year 2000 $\Delta V \leq 0.2$ mag (with the exception of some 2007 photometric data) confirming that the spectral type is B7 and supporting the contention that only large amplitude photometric variabilities are present for spectral type B8 or later.} 
         \label{fig:deltav}
   \end{figure}

With the decades of photometric measurements of W90 and the alleged spectral type evolution discussed in the previous section, we tested a correlation advanced by Finkenzeller \& Mundt (1984) and Bibo \& Th\'e (1990), and confirmed by van den Ancker et al. (1998) and Herbst \& Shevchenko (1999).  Herbst \& Shevchenko (1999), using a large large data set of 230 UX Ori objects which included W90 observations, established that large amplitude photometric variability ($\Delta V \geq 0.2$ mag) is present exclusively in stars with spectral type B8 or later. If this empirical correlation holds, we should not expect photometric variabilities larger than 0.2 mag, and we could estimate the crossing time of the spectral type to B7.  For the photometric data, $V$, of W90 we formed the index $\Delta V$ for the years with multiple observations by subtracting $V_{min}$ from $V_{max}$ for each individual year. The index $\Delta V$ is displayed in the bottom panel of Fig.~\ref{fig:deltav}, showing that some time after the year 2000 the variabilities in $V$ are the smallest in the whole photometric history of W90, with the possible exception of the 2007 photometric data. This analysis supports both the validity of the correlation and the recent spectral type determination of B7. In summary, the apparent changes toward earlier spectral types, mimicking a hotter photosphere with time, are due to the increasing visible accretion temperature that is being sampled in the line-of-sight.   

\subsection{Anomalous extinction} 

Although UV wavelengths are subjected to differential extinction, they have the advantage of being relatively free of emission lines compared with infrared wavelengths. Accurate flux measurements in the far-UV could help us to obtain more precise R$_v$ values through the calibration described by Cardelli \& Clayton (1988). Previous studies on the UV extinction in young open clusters (c.f., Massa \& Savage 1985, Massa \& Fitzpatrick 1986) have indicated that 
the differences of structure from star-to-star are within the observational errors, suggesting that open clusters such as NGC 2244 have a unique non-variable UV extinction curve. 

Earlier UV observations of W90 by Sitko et al. (1984) indicated that the grain size was larger than those in the normal interstellar medium and that the graphite-silicate mixture seems to fit the extinction curve, although this is noted as highly uncertain. It was also suggested that a larger value of R$_v$ ($\geq$ 7) was required to deredden W90 to its proper position on the color-color diagram. 

By analyzing polarization data, one can also obtain insights into the extinction characteristics. Intrinsic polarization for this star has also been found to be variable by Breger (1974). He found a local maximum in 1973 at $\lambda _{max}$=0.65 $\mu$m. Using the formula $$R_v=(5.6\pm 0.3)\ast \lambda_{max} ,$$ we can derive a value of R$_v$=3.64 which seems to be a lower limit for its extinction correction. P\'{e}rez et al. (1987) derived a value of R$_v$=5.2 by fitting the unreddened spectral energy distribution (SED) with the best Kurucz model (T=12,000 K, log g=3.5). Recently, Hern\'andez et al. (2004), following the suggestion made by Testi et al. (1998), demonstrated for a larger statistical sample of 55 HAeBe stars(including W90) that a value of R$_v$=5.0 was a better extinction correction for the heavily reddened objects in their sample, suggesting a larger grain size in their circumstellar environments. It is interesting to note that by using larger R$_v$ values to estimate stellar parameters, objects like W90 appear much younger ($t_{age} \leq$ 0.2 Myr; Testi et al. 1998).  

\subsection{The spectral energy distribution}

\begin{figure}
   \centering
 \includegraphics[angle=-90,width=8.5cm]{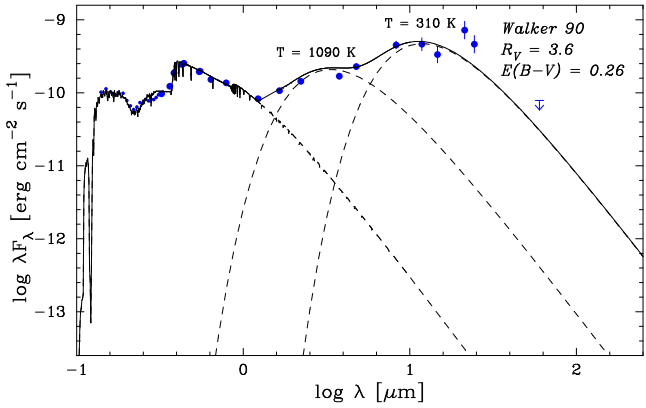}
      \caption{Spectral energy distribution derived for W90 by using an abnormal extinction law of R$_v$=3.6. Note the onset of a large IR excess starting around 1.25 $\mu$m.  The effective temperature derived corresponds to a B7--8V star. Two Planck emission functions having temperatures of 1090 K and 310 K are fitted to the near- and far-infrared excesses.}
         \label{fig:sed}
   \end{figure}

Using mean observed photometric values, we have derived a SED by using R$_{v}$=3.6. The fit of the photospheric temperature of T$_{eff}$=12,000 K, which corresponds to a B7--8V star, is somewhat satisfactory in the blue and UV colors. We note the large IR excess starting at 1.25 $\mu$m. The SED is presented in Fig.~\ref{fig:sed}. We do not detect the optical depression of about 3 magnitudes at $V$ noted by Rydgren \& Vrba (1987).  
By modeling the infrared excess of this star we were able to fit two dust shells having temperatures of 1,090 K and 310 K, respectively, illustrated also in Fig.~\ref{fig:sed}. This result adds strong support to the thermal re-emission argument as the responsible mechanism of the infrared excess. 

\subsection{Inconclusive x-ray emission}

Simon et al. (1985) tagged W90 as an {\it Einstein} X-ray emitter candidate indicating that the thick disk-like shell around this star could be fairly transparent to high energy photons.  However, we note that two surveys of x-ray sources in NGC 2264 by Flaccomio et al. (2000) of {\it ROSAT HRI} observations (average of log$(L_X) \sim 30.49 \pm 0.03$ for the detected sample, where $L_X$ is in $ergs~ s^{-1}$) and by Ram\'{\i}rez et al. (2004) based on {\it ACIS-Chandra} observations failed to detect W90 to the upper limit of log$(L_X) \sim 28.48$.  Hamaguchi et al. (2005), doing an {\it ASCA} archival survey, also failed to detect W90 to their exposure limit of log$(L_X) \sim 32.0$. They also provided a revised {\it ROSAT HRI} upper limit detection for W90 of log$(L_X) = 31.3.$  In analyzing archival {\it Chandra} data, Stelzer et al. (2006) examined the role of previously unseen late-type companions in the sample and confirmed the non-detection of W90 with an upper limit of log$(L_X) \sim 28.8$, in close agreement with the earlier result by Ram\'{\i}rez et al. (2004).

Most recent x-ray surveys by Flaccomio et al. (2006), based on 100ks long of {\it ACIS-Chandra} data and Dahm et al. (2007) of {\it XMM-Newton EPIC} observations of NGC 2264, failed to detect W90 within their sensitivities. Although the decline in x-ray emission for Herbig Ae/Be stars with increasing age is well documented (e.g., Hamaguchi et al. 2005), suggesting an age for W90 of several million years, as $L_X$ decreases below $10^{30}~ erg~ s^{-1}$.  We further discuss the age of W90 in the following section, based on the observed disk size.  

\subsection{Upper limit to disk size}

In both infrared images secured with the VLT {\it VISIR} detector, W90 was indistinguishable from a point-source. From these data, we derive an upper limit of $<$ 0.1" for the diameter of the emitting region in the thermal infrared. Assuming the most common distance adopted toward NGC 2264 of 800 pc (e.g., Dahm \& Simon 2005, Teixeira et al. 2006), an upper limit of 80 AU is derived for the outer disk size of an optically thick disk.  This limit fits in the low-end of other disk surveys of HAeBe stars, such as the one conducted by Eisner et al. (2004), who analyzed 24 HAeBe objects at 2.2~$\mu$m and found disk sizes ranging from 30 to 400 AU. Moreover, W90 follows the low end trend described by Leinert et al. (2004) of the [12-15] {\it IRAS} colors and the disk sizes. 

This upper limit for the disk size can be further analyzed to get an estimate of the stellar and disk system age. Dent et al. (2005) suggested that disk sizes seem to decrease with time, and for an 80 AU disk system a representative age of 7--10 Myr is empirically assigned. This range fits well within the upper end of the mean ages for the overall cluster, NGC 2264, as compiled by Dahm \& Simon (2005) spanning from 0.1 to 10 Myr. 

\subsection{Spectral features}

\subsubsection{Inverse P-Cygni profiles in Balmer lines} 

The H$\alpha$ (6562 \AA, Fig.~\ref{fig:halpha}) and H$\beta$ (4861 \AA, Fig.~\ref{fig:hbeta}) lines are in emission with a complex underlying structure of at least 3--4 individual components. H$\alpha$ spectra revealed strong emission profile variability with a broadening feature of $\pm 400$ km/s plus an inverse P-Cygni (IPC) feature of $W_{H\alpha} \geq 10$\AA. H$\gamma$ and H$\delta$ (Fig.~\ref{fig:hdelta}) are in absorption, showing asymmetric and rotationally broadened profiles. Higher order Balmer lines starting with H$\epsilon$ through H8-H15 are all present in absorption with the same asymmetric structure shown by earlier members of the series. 

From Schmidt film H$\alpha$, Reipurth et al. (2004) reported that W90 had an average H$\alpha$ emission.  H$\alpha$ profiles (Fig.~\ref{fig:halpha}) observed in W90 are clearly different from all of the H$\alpha$ profiles which Reipurth et al. (1996) proposed as typical for characterizing pre-main sequence objects. Variable H$\alpha$ emission and IPC profiles were observed in just two of 18 HAeBe stars (UX Ori and BF Ori) and three of the 43 T Tauri stars (SZ Cha, SZ 82, and AS207) by Reipurth et al. (1996). Vieira et al. (2003) found similar behavior in two HAeBe candidates (PDS 018 and PDS 024). However, both the H$\alpha$ emission strength and IPC variability observed in W90 by us were significantly more extreme than what was seen in the examples of Reipurth et al. (1996) and Viera et al. (2003). The IPC profiles in W90 are not limited to the low Balmer lines.  In fact, Dahm \& Simon (2005), via IRTF measurements, found IPC profiles for He I (1.0833, 2.0587 $\mu$m). In addition, they detected the Brackett series and many of the Paschen lines in emission as well as several Fe II and [Fe II] lines.

Unlike P-Cygni profiles, which are attributed to strong stellar winds and mass loss, IPC features are rare and widely interpreted as evidence of mass infall (e.g., Hartmann et al. 1994, Edwards et al. 1994). Li \& Rector (2004) argued, alternatively, that H$\alpha$ IPC could be attributed to a high-inclination angle of the edge-on relic disk. Rapid changes of the blue and red peaks in the IPC profile, such as we observed in W90, could be caused by the presence of a variable structure and clumpiness in the circumstellar disk (de Winter et al. 1999). Furthermore, there appears to be a correlation between UX Ori objects and the presence of H$\alpha$ IPC profiles (Grinin \& Rostopchina 1996), relationship which is reinforced by our data since W90 has some photometric resemblance to UX Ori objects (Fig.~\ref{fig:color var}). In summary, the strong presence of IPC profiles in several low order Balmer lines at multiple epochs over a period of months, not only suggest prolonged and thus substantial mass infall, but is not consistent with the behavior of T Tauri stars, which display fluctuations from mass infall to mass over such time scales. Thus the behavior of the low order Balmer lines in W90 is suggestive of it being a high-mass object.

\subsubsection{Optical and near-IR lines: Fe, Mg, Na, He, Ca, O, Ca II and Paschen lines}

The metal lines of Fe I and II are mostly absent, with the possible exception of Fe I (4325 \AA), which appears weakly in emission. Mg I lines (4481, 4703, 6318, etc.) are equally missing from the spectra.  

The Na I doublet at 5890, 5896 \AA~ is in emission, showing remarkably narrow profiles characteristic of an H II region.  He lines (4026, 4121, 4144, 4471, and 5875 \AA) are weakly visible in broad absorptions. This suggests that the spectral type of W90 is probably later than B9. But the Ca II H and K (3934 and 3968 \AA) lines, as shown in Fig.~\ref{fig:caiik}, are conspicuously part of a broad absorption, suggesting that the spectral type of the material sampled is not later than A0. 

Ca II triplet and Paschen lines are both in emission and absorption.  Ca II lines in emission have been suggested by Catala et al. (1986) and Hamann \& Persson (1992) to be a unique characteristic of HAeBe stars. 

In summary, the overall multiwavelength spectrum of W90 does not appear to be dominated by a stellar photosphere. Rather, the spectra are a superposition of an emission nebulosity, a dense HII region, and an accretion disk, which points to the complexity of the immediate environment of this object.

\subsubsection{Ultraviolet continuum and lines}

Previous UV data analyses of W90 were published by Sitko et al. (1984), Simon et al. (1985), and Valenti et al. (2000, 2003).  Sitko et al. (1984) used the {\it IUE} data to study the anomalous dust extinction toward this object, arguing that a larger size graphite-silicate mixture can explain it. However, they did not include any line identification in the spectra. From {\it IUE} spectra of 19 stars in NGC 2264, Simon et al. (1985) found a far more diverse star sample compared with clusters previously studied. They detected no chromospheric emission for stars brighter than 12 magnitudes that lie above the ZAMS. We note here that the chromospheric emission interpretation was an early attempt to explain emission phenomena likely arising from more energetic processes, such as accretion flows onto the central star. Even for stars with later spectral types that allow chromospheres to exist, such as T Tauri stars, this interpretation has been out of favor (e.g., Johns-Krull et al. 2000). Furthermore, Simon et al. (1985) expressed some uncertainties as to whether the circumstellar and interstellar extinction masked any stellar emission feature.

A rising continuum flux toward shorter wavelengths was detected (Fig.~\ref{fig:swp}, Fig.~\ref{fig:swp_color}). Although this was interpreted as evidence of an early spectral type by Imhoff \& Appenzeller (1987), we believe that the rising UV continuum of W90 probably has a non-photospheric origin.  In the collection of SWP spectra of 74 HAeBe stars (Valenti et al. 2000, their Fig. 4), W90 shows a distinctive rising continuum with some weak emissions superimposed, similar to the photospheric flux seen in earlier spectral type objects among their sample.  

Although calibration errors can be reduced by co-adding {\it IUE} spectra obtained at different epochs Valenti et al. (2003), we did not attempt to co-add the data for line identification due to significant flux variations (beyond the reproducibility errors of the {\it IUE} cameras $\sim$ 3\%), which are clearly present among the different spectra. A relative amplitude of fluctuations of 5.9\% was measured for the SWP camera (Valenti et al. 2000) and also was measured for the LWP cameras (Valenti et al. 2003). We tentatively detect the N V emission line (1240 \AA), which arises in a super-ionized region (T$\sim 2\times 10^5$~K). In a study of 49 T Tauri stars, Johns-Krull et al. (2000) concluded that the SWP continuum fluxes in the classical T Tauri stars appear to originate in a $\sim$ 10,000 K optically thick plasma.  By analogy, we suggest that the steeper continuum flux in W90 could originate from the hotter, $\sim 10^{5}$ K plasma, which is implied by the N V line.  Besides this line, the other line identification which we could make with confidence is the C IV absorption (1550 \AA), which is typically a strong feature in hot stars.

In Fig.~\ref{fig:lwp}, we show the LW spectra available (excluding LWP 27428), where it is possible to detect several incipient emission lines such as Fe I, Fe II, Mg II, etc. The presence of Fe lines and the Mg II lines in emission are excellent diagnostics of temperatures and significant densities for the environment surrounding W90. Many other emission lines appear blended (e.g., Cr II, Ni III, Mn II, etc.) so their identifications are somewhat unreliable in these spectra. W90 follows the behavior described by Imhoff (1994) for pre-main sequence objects with Mg II in emission, which also present strong near-IR excesses. Despite the flux variations, a mean deficiency can be seen around 2200~\AA~ (Fig.~\ref{fig:lwp_color}, bottom panel). Such a dip has commonly been interpreted as arising from a dense line-of-sight due to interstellar extinction of small grains.  

\section{Summary and conclusions}

The complex and bizarre behavior of W90 has been demonstrated and documented with decades of multiwavelength measurements from new and archival data.  First, we summarize our most certain conclusions as interpreted and derived primarily from observations discussed here. Second, we conclude with some exploratory insights on the physical conditions of W90 and future research directions. 
\begin{enumerate}
\item Our recent photometric data in 2007 indicate that W90 is at its brightest recorded optical magnitude ($\overline{12.47} \pm 0.06$).  Individual data points in 2007 are even brighter than the 1971 photometric $V$ levels. We note that from 2004 to 2005, the annual average of $V$ changes by -0.06, which is the same value from 2005 to 2006, however, from 2006 to 2007, $V$ decreases by 0.19. The brightening of W90 is accompanied with less dispersion in the $V$ values ($\Delta V\leq 0.20$), supporting a current spectral type of B7. The aperiodic long-term photometric variability shown in Fig.~\ref{fig:long-term var} is rather common for young objects and has rarely been accurately measured in other similar objects, with the exception of the late-type star KH 15D in NGC 2264 (Herbst et al. 2002). However, long-term photometric variability in KH 15D (short-term variability is due to a well-documented eclipse) has been attributed to the precession of its circumstellar disk. 

\item The sustained and well-documented evolution in spectral type (from A2/A3 to currently B7; however, it has been classified as early as B4) in the last five decades, the slight decrease of the near-infrared K fluxes, IPC in low order Balmer lines, and the irregular photometric variability during this period suggest that W90 is surrounded by an optically thick accretion disk. This means that this disk is diminishing its overall volume in the line-of-sight, or is becoming less dense, or appears tilting its angle. A primary thick disk could subsequently explain polarization levels, abnormal extinction, spectral and photometric variability, the incipient blueing effect, and the presence of the emission lines discussed. 

\item The dimension of the surrounding disk is certainly constrained by our detection of a point source at infrared wavelengths no larger than 80 AU in diameter. Following Dent et al. (2005), this observation also imposes a stellar age range of several million years, which fits the upper end of the cluster age as determined by Dahm \& Simon (2005). Likewise, the modeling of the IR excess suggests the existence of two dust shells with temperature components of 1,090 K and 310 K.  

\item The steep UV continuum fluxes (mimicking a star as early as B4) and the tentative identification of N V emission (implying the existence of plasma temperatures of T$\sim 2\times 10^5$~K) unambiguously demonstrate the detection of a non-stellar energetic component, such as the flux arising from a thermally inhomogeneous accretion disk.   

\item The obscuring disk or cavity surrounding W90 appears opaque to radio emission as reported by Fuente et al. (1998), since no mm- or cm-continuum emissions were detected. Similarly, speckle observations by Leinert et al. (1997) to detect a binary system were reported negative.  Nevertheless, the detected rising UV continuum finds ways to escape suggesting that it is either produced in the outer layers of the disk or there is a favorable viewing angle into the accretion flow. We suggest that the accretion disk alignment is such that we are able to probe the strong UV continuum arising from collisionally excited material in the disk. 

\item It is likely that W90 is a flared disk system, as suggested by Acke \& van den Ancker (2006) through the detection of a tentative PAH emission at 3.3 $\mu$m, which is common in flared disk systems. Similarly, Dietrich (2007) detected in W90 strong PAH emissions at 6.2 and 11.2 $\mu$m and weaker emissions at 7.7 and 8.6 $\mu$m. He also found a strong 10 $\mu$m silicate feature and weak emission bands at 24, 28, 34 $\mu$m, probably caused by crystalline silicates.  Moreover, Acke \& van den Ancker (2006) reported no detection of the nearby nanodiamond features at 3.43 and 3.53 $\mu$m, observational evidence determined to be rare among HAeBe stars. Due to the strong UV continuum present in W90 and the 2200~\AA~ absorption bump, one expects a strong PAH emission as predicted by Meeus et al. (2001) if we consider that PAH molecules are excited by UV stellar radiation. Instead, this result confirms the assertion made by Acke \& van den Ancker (2004) that the existence of a self-shadowed inner rim of the disks prevents the PAH from being excited by the central star, yielding weak or non-existent PAH emissions at 3.3 $\mu$m.  
\end{enumerate}

W90, as a member of the BMS group in NGC 2264 recently described by Sung et al. (2008), is perhaps seen in light scattered through material in the circumstellar disk. The seeming long-term clearing of the disk or surroundings of W90, as detected by the changes in optical, near- infrared photometry and spectral types presented in our study, has certainly not been sufficient to let high and medium energy photons escape, namely X-ray and radio emission. But the unambiguous detection of highly variable IPC features in many spectral lines, such as the Balmer lines, clearly demonstrates the evidence of some infall process, which we believe is the manifestation of accretion flows within favorable viewing geometry.  If this phenomenological picture of W90 is correct, we anticipate continuous and incremental spectral, photometric, and, in general, emission changes in the upcoming decades. Of the physical mechanisms and interpretations outlined here, such as UXOR-like variability, accretion flows, disk precession, etc., we do not favor one over the other, since our physical picture of this object is still incomplete. 

The remarkable observational uniqueness of W90 is not based on a specific measurement but is the aggregate value of comparing and analyzing the large volume of multiwavelength evidence across several decades.  By using all the data presented here plus new observations, we expect to pursue some modeling of the complex behavior of W90. However, this is clearly beyond the scope of the present study.                                   
\begin{acknowledgements}
      This research has made use of the Simbad database, operated at CDS, Strasbourg, France. This work is based in part on observations made with the {\it Spitzer} Space Telescope operated by the Jet Propulsion Laboratory, California Institute of Technology, under a contract with NASA.
This publication makes use of data products from the Two Micron All Sky Survey, which is a joint project of the University of Massachusetts and the Infrared Processing and Analysis Center/California Institute of Technology, funded by the National Aeronautics and Space Administration and the National Science Foundation. We thank the Palomar Observatory for the generous assignment of time for the Echelle observations. We thank Dr. Derck Massa for providing the color figures (Figs.~\ref{fig:swp_color} \& \ref{fig:lwp_color}), which utilize his {\it IUE} NEWSIPS flux calibration corrections.  We also acknowledge the use of the SAAO facilities and thanks to Dr. C. D. Laney for some of the IR observations and $BVR_{c}I_{c}$ photometry of standard stars in the W90 field. We finally thank the support by the U.S. Department of Energy, through the Los Alamos Laboratory-Directed Research and Development (LDRD 20080085DR) funds, used to complete the analysis work included in this paper (LA-UR-08-2165). We also thank the anonymous referee for his comments which helped us to improve the presentation and to clarify key points.  
\end{acknowledgements}
  

\begin{thebibliography}{}

\bibitem[Acke \& van den Ancker(2006)]{2006A&A...457..171A} Acke, B., \& 
van den Ancker, M.~E.\ 2006, \aap, 457, 171

\bibitem[Acke \& van den Ancker(2004)]{2004A&A...426..151A} Acke, B., \& 
van den Ancker, M.~E.\ 2004, \aap, 426, 151 

\bibitem[Allen(1973)]{1973MNRAS.161..145A} Allen, D.~A.\ 1973, \mnras, 161, 
145 

\bibitem[1979]{} Bessell, M. S. 1979, PASP, 91, 589

\bibitem[Bibo \& The(1990)]{1990A&A...236..155B} Bibo, E.~A., \& Th\'{e}, 
P.~S.\ 1990, \aap, 236, 155 

\bibitem[1992]{Bhatt} Bhatt, H.C., \& Sagar, R. 1992, A\&AS, 92, 473

\bibitem[1974]{}Breger, M. 1974, ApJ, 188, 53

\bibitem[Breger(1972)]{1972ApJ...171..539B} Breger, M.\ 1972, \apj, 171, 
539 

\bibitem[1988]{}Cardelli, J. A., \& Clayton, G. C. 1988, AJ, 95, 516 

\bibitem[1986]{}Catala, C., Czarny, J., Felenbok, P., \& Praderie, F. 1986, A\&A, 154, 103

\bibitem[Crawford \& Mander(1966)]{1966AJ.....71..114C} Crawford, D.~L., \& 
Mander, J.\ 1966, \aj, 71, 114

\bibitem[2007]{Dahm07} Dahm, S.E., Simon, T., Proszkow, E.M., \& Patten, B.M. 2007, AJ, 134, 999

\bibitem[2005]{Dahm05} Dahm, S.E., \& Simon, T. 2005, AJ, 129, 829

\bibitem[Dent et al.(2005)]{2005MNRAS.359..663D} Dent, W.~R.~F., Greaves, 
J.~S., \& Coulson, I.~M.\ 2005, \mnras, 359, 663 

\bibitem[de Winter et al.(2001)]{2001A&A...380..609D} de Winter, D., van 
den Ancker, M.~E., Maira, A., Th{\'e}, P.~S., Tjin A Djie, H.~R.~E., 
Redondo, I., Eiroa, C., \& Molster, F.~J.\ 2001, \aap, 380, 609 

\bibitem[de Winter et al.(1999)]{1999A&A...343..137D} de Winter, D., Grady, 
C.~A., van den Ancker, M.~E., P{\'e}rez, M.~R., \& Eiroa, C.\ 1999, \aap, 
343, 137 

\bibitem[]{}Dietrich, H. 2007, Ph.D. Thesis, Faculty of Physics and Astronomy, Univ. of Heidelberg, p. 64

\bibitem[Edwards et al.(1994)]{1994AJ....108.1056E} Edwards, S., Hartigan, 
P., Ghandour, L., \& Andrulis, C.\ 1994, \aj, 108, 1056 

\bibitem[Eisner et al.(2004)]{2004ApJ...613.1049E} Eisner, J.~A., Lane, 
B.~F., Hillenbrand, L.~A., Akeson, R.~L., \& Sargent, A.~I.\ 2004, \apj, 
613, 1049

\bibitem[Fazio, G.G. (2004)]{faz04} Fazio, G.G., et al.  2004, ApJ, 154, 10

\bibitem[Finkenzeller \& Mundt(1984)]{1984A&AS...55..109F} Finkenzeller, 
U., \& Mundt, R.\ 1984, \aaps, 55, 109 

\bibitem[Flaccomio et al.(2006)]{2006A&A...455..903F} Flaccomio, E., 
Micela, G., \& Sciortino, S.\ 2006, \aap, 455, 903 
 
\bibitem[2000]{} Flaccomio, E. et al. 2000, A\&A, 355, 651

\bibitem[Fuente et al.(1998)]{1998A&A...334..253F} Fuente, A., 
Martin-Pintado, J., Bachiller, R., Neri, R., \& Palla, F.\ 1998, \aap, 334, 
253 

\bibitem[1994]{} Ghandour, L. et al. 1994, ASP Conf. Ser. 62, 223

\bibitem[1994]{} Grinin, V.P., Th\'e, P.S., de Winter, D., et al. 1994, A\&A, 292, 165
 
\bibitem[Grinin \& Rostopchina(1996)]{1996ARep...40..171G} Grinin, V.~P., 
\& Rostopchina, A.~N.\ 1996, Astronomy Reports, 40, 171 

\bibitem[Hamaguchi et al.(2005)]{2005ApJ...618..360H} Hamaguchi, K., 
Yamauchi, S., \& Koyama, K.\ 2005, \apj, 618, 360 

\bibitem[1992]{} Hamann, F., \& Persson, S.E. 1992, ApJS, 82, 285

\bibitem[Hartmann et al.(1994)]{1994ApJ...426..669H} Hartmann, L., Hewett, 
R., \& Calvet, N.\ 1994, \apj, 426, 669 

\bibitem[1988]{} Herbig, G. H., \& Bell, K. R. 1988, ``Third Catalog of Emission-Line Stars of the Orion Population,'' Lick Observatory Bulletin No. 1111

\bibitem[1972]{}Herbig, G. H., \& Rao, N. K. 1972, ApJ, 174, 401

\bibitem[1960]{} Herbig, G. H. 1960, ApJS, 4, 337

\bibitem[1954]{} Herbig, G. H. 1954, ApJ, 119, 483

\bibitem[Herbst et al.(2002)]{2002PASP..114.1167H} Herbst, W., et al.\ 
2002, \pasp, 114, 1167 

\bibitem[Herbst \& Shevchenko(1999)]{1999AJ....118.1043H} Herbst, W., \& 
Shevchenko, V.~S.\ 1999, \aj, 118, 1043 

\bibitem[Herbst et al.(1994)]{1994AJ....108.1906H} Herbst, W., Herbst, 
D.~K., Grossman, E.~J., \& Weinstein, D.\ 1994, \aj, 108, 1906 

\bibitem[2004]{} Hern\'andez, J., Calvet, N., Brice\~no, C., Hartmann, \& L., Berlind, P. 2004, AJ, 127, 1682

\bibitem[]{}Hillenbrand, L.~A. 1995, Ph.D. Thesis, University of Massachusetts

\bibitem[Hillenbrand et al.(1992)]{1992ApJ...397..613H} Hillenbrand, L.~A., 
Strom, S.~E., Vrba, F.~J., \& Keene, J.\ 1992, \apj, 397, 613 

\bibitem[Imhoff(1994)]{1994ASPC...62..107I} Imhoff, C.~L.\ 1994, The Nature 
and Evolutionary Status of Herbig Ae/Be Stars, 62, 107

\bibitem[Imhoff \& Appenzeller(1987)]{1987euwi.conf..295I} Imhoff, C.~L., 
\& Appenzeller, I.\ 1987, Exploring the universe with the IUE satellite, 
p.~295 - 319

\bibitem[Johns-Krull et al.(2000)]{2000ApJ...539..815J} Johns-Krull, C.~M., 
Valenti, J.~A., \& Linsky, J.~L.\ 2000, \apj, 539, 815 

\bibitem[Johnson(1966)]{1966ARA&A...4..193J} Johnson, H.~L.\ 1966, \araa, 
4, 193 

\bibitem[Kwon \& Lee(1983)]{1983JKAS...16....7K} Kwon, S.~M., \& Lee, 
S.-W.\ 1983, Journal of Korean Astronomical Society, 16, 7 

\bibitem{}Lagage P.O., et al., 2004, The Messenger 117, 12

\bibitem[Lamm et al.(2004)]{2004A&A...417..557L} Lamm, M.~H., Bailer-Jones, 
C.~A.~L., Mundt, R., Herbst, W., \& Scholz, A.\ 2004, \aap, 417, 557

\bibitem[Leinert et al.(2004)]{2004A&A...423..537L} Leinert, C., et al.\ 
2004, \aap, 423, 537 

\bibitem[Leinert et al.(1997)]{1997A&A...318..472L} Leinert, C., Richichi, 
A., \& Haas, M.\ 1997, \aap, 318, 472

\bibitem[Li \& Rector(2004)]{2004ApJ...600L..67L} Li, J.~Z., \& Rector, 
T.~A.\ 2004, \apjl, 600, L67 

\bibitem[Massa \& Fitzpatrick(2000)]{2000ApJS..126..517M} Massa, D., \& 
Fitzpatrick, E.~L.\ 2000, \apjs, 126, 517 

\bibitem[1986]{}Massa, D., \& Fitzpatrick, E. L. 1986, ApJS, 60, 305

\bibitem[1985]{} Massa, D., \& Savage, B. D. 1985, ApJ, 299, 905

\bibitem[Meeus et al.(2001)]{2001A&A...365..476M} Meeus, G., Waters, 
L.~B.~F.~M., Bouwman, J., van den Ancker, M.~E., Waelkens, C., \& Malfait, 
K.\ 2001, \aap, 365, 476 

\bibitem[Mendoza V.~\& Gomez(1980)]{1980MNRAS.190..623M} Mendoza V., E.~E., 
\& Gomez, T.\ 1980, \mnras, 190, 623 

\bibitem[Neri et al.(1993)]{1993A&AS..102..201N} Neri, L.~J., Chavarria-K., 
C., \& de Lara, E.\ 1993, \aaps, 102, 201 

\bibitem[1992]{} P\'{e}rez, M. R., Geisler, D., Joner, M., \& de Winter, D., 1992, IAU Circular, No. 5486

\bibitem[1991]{}P\'{e}rez, M. R., \& Loomis, C. 1991, Record of the IUE Three Agency Coordination Meeting (NASA/ESA/SERC), Nov 19–21, 1991, GSFC, F-13

\bibitem[1987]{} P\'{e}rez, M. R., Th\'{e}, P. S., \& Westerlund, B. E. 1987, PASP, 99, 1050 

\bibitem[2004]{} Ram\'{\i}rez, S.V. et al. 2004, AJ, 127, 2659
 
\bibitem[Reipurth et al.(2004)]{2004AJ....127.1117R} Reipurth, B., Pettersson, B., Armond, T., Bally, J., \& Vaz, L.~P.~R.\ 2004, \aj, 127, 1117

\bibitem[]{}Reipurth, B., Pedrosa, A., \& Lago, M.T.V.T. 1996, A\&AS, 120, 229 

\bibitem[Rieke et al. (2004)]{rie04} Rieke, G. et al. 2004, ApJS 154, 25

\bibitem[1987]{}Rydgren, A. E., Vrba, F. J. 1987, PASP, 99, 482

\bibitem[Safier(1995)]{1995ApJ...444..818S} Safier, P.~N.\ 1995, \apj, 444, 
818

\bibitem[Sagar \& Joshi(1983)]{1983MNRAS.205..747S} Sagar, R., \& Joshi, 
U.~C.\ 1983, \mnras, 205, 747 

\bibitem[1985]{Simon} Simon, T., Herbig G., \& Boesgaard, A.M. 1985, ApJ, 293, 542

\bibitem[1984]{}Sitko M.L., Simon, T., \& Meade, M.R. 1984, PASP, 96, 54

\bibitem[Steenman \& The(1991)]{1991Ap&SS.184....9S} Steenman, H., \& Th\'{e}, P.~S.\ 1991, \apss, 184, 9 

\bibitem[Stelzer et al.(2006)]{2006A&A...457..223S} Stelzer, B., Micela, 
G., Hamaguchi, K., \& Schmitt, J.~H.~M.~M.\ 2006, \aap, 457, 223 

\bibitem[Strom et al.(1972)]{1972ApJ...171..267S} Strom, S.~E., Strom, 
K.~M., Brooke, A.~L., Bregman, J., \& Yost, J.\ 1972, \apj, 171, 267

\bibitem[Sung et al.(2008)]{2008AJ....135..441S} Sung, H., Bessell, M.~S., 
Chun, M.-Y., Karimov, R., \& Ibrahimov, M.\ 2008, \aj, 135, 441 

\bibitem[Sung et al.(1997)]{1997AJ....114.2644S} Sung, H., Bessell, M.~S., 
\& Lee, S.-W.\ 1997, \aj, 114, 2644 

\bibitem[Teixeira et al.(2006)]{2006ApJ...636L..45T} Teixeira, P.~S., et 
al.\ 2006, \apjl, 636, L45 

\bibitem[Testi et al.(1998)]{1998A&AS..133...81T} Testi, L., Palla, F., \& 
Natta, A.\ 1998, \aaps, 133, 81

\bibitem{}van den Ancker M.E., de Winter D., Tjin A Djie H.R.E., 1998, A\&A 330, 145

\bibitem[Valenti et al.(2003)]{2003ApJS..147..305V} Valenti, J.~A., Fallon, 
A.~A., \& Johns-Krull, C.~M.\ 2003, \apjs, 147, 305 

\bibitem[1965]{}Vasilevskis, S., Sanders, W.L., \& Balz, Jr., A.G. 1965, AJ, 70, 797

\bibitem[]{}Vieira, S.L.A., Corradi, W.J.B., \& Alencar, S.H.P. et al., 2006, AJ, 126, 2971

\bibitem[1956]{}Walker, M. F. 1956, ApJS, 2, 365

\bibitem[Warner et al.(1977)]{1977ApJ...213..427W} Warner, J.~W., Strom, 
S.~E., \& Strom, K.~M.\ 1977, \apj, 213, 427 

\bibitem[1978]{}Young, A. 1978, PASP, 90, 144

\bibitem[1984]{} Zhao et al. 1984. Special Issue for Tables of Membership for 42 Open Clusters, Academia Sinica, Shangai, p. 213.
\end{thebibliography}
\end{document}